\documentclass[aps,prb,twocolumn,showpacs,groupaddress,floatfix]{revtex4}
\usepackage{graphicx,graphics}
\usepackage{dcolumn}
\usepackage{amsmath,amssymb,amsfonts}
\usepackage{latexsym,verbatim}
\usepackage{bm}
\usepackage{color}
\usepackage{ulem}
\usepackage[breaklinks=true,colorlinks,citecolor=blue,linkcolor=blue,urlcolor=blue]{hyperref}

\def\be{\begin{equation}}
\def\ee{\end{equation}}
\def\Lambdabold{\mbox{$\bm \Lambda$}}
\begin{document}
\title{Violation of the Wiedemann-Franz law in clean graphene layers}

\author{Alessandro Principi}
\email{principia@missouri.edu}
\affiliation{Department of Physics and Astronomy, University of Missouri, Columbia, Missouri 65211, USA}	
\author{Giovanni Vignale}
\affiliation{Department of Physics and Astronomy, University of Missouri, Columbia, Missouri 65211, USA}

\begin{abstract}
The Wiedemann-Franz law,  connecting the electronic thermal conductivity to the electrical conductivity of a disordered metal, is generally found to be well satisfied even when  electron-electron (e-e) interactions are strong.  In ultra-clean conductors, however, large deviations from the standard form of the law are expected, due to the fact that e-e interactions affect the two conductivities in radically different ways.   Thus, the standard Wiedemann-Franz ratio between the thermal and the electric conductivity is reduced by a factor $1+\tau/\tau_{\rm th}^{\rm ee}$, where $1/\tau$ is the momentum relaxation rate, and $1/\tau_{\rm th}^{\rm ee}$ is the  relaxation time of the thermal current due to e-e collisions.   Here we study the density and temperature dependence of $1/\tau_{\rm th}^{\rm ee}$ in the important case of doped, clean single layers of graphene, which exhibit record-high thermal conductivities. We show that at low temperature  $1/\tau_{\rm th}^{\rm ee}$  is $8/5$ of the quasiparticle decay rate.  We also show that the many-body renormalization of the thermal Drude weight coincides with that of the Fermi velocity. 
\end{abstract}
%\pacs{73.20.Mf,71.45.Gm,78.67.Wj}
%
\maketitle

\section{Introduction}
Thermoelectric phenomena, in which heat is converted to electric power and viceversa, have long been the subject of intense research activity. In recent years, theoretical interest in these phenomena has reached new heights, due to their implications for the development of sustainable energy sources~\cite{Nolas_book,Dubi_rmp_2011}. Understanding thermoelectric phenomena in semiconductors is also extremely important to try and reduce heat losses in electronic circuits. 
In this respect graphene, a monolayer of carbon atoms packed in a two-dimensional (2D) honeycomb lattice~\cite{castroneto_rmp_2009,kotov_rmp_2012}, is a potentially important material~\cite{Balandin_nanolett_2008,Wei_prl_2009}. Experiments have shown~\cite{Balandin_nanolett_2008} that the thermal conductivity of pristine suspended graphene (due to phonons) has an extremely high value, as compared with other semiconductors. Such a large thermal conductivity may be beneficial for electronic applications and thermal management~\cite{Balandin_nanolett_2008}.

In this Letter we focus on the thermal current, precisely defined as the temperature $T$ times the entropy current carried by a high-mobility electron gas in a layer of doped graphene. It is well known that the electronic thermal ($\sigma_{\rm th}$) and charge ($\sigma_{\rm c}$) d.c. conductivities of such a gas are connected by the  Wiedemann-Franz (WF) law~\cite{Trushin_prl_2007,Wiedemann_Franz}, which states that 
\begin{equation} \label{eq:conductivity_ratio}
\frac{\sigma_{\rm th}}{\sigma_{\rm c} T} = \frac{\pi^2 k_{\rm B}^2}{3 e^2} 
~,
\end{equation}
where the quantity on the right hand side -- the so-called ``Lorenz number" --  is a universal constant, independent of material parameters. This elegant statement reflects the fact that a single set of carriers (electrons) transport both the charge and the thermal energy, and that the scattering mechanism (mainly electron-impurity scattering at low temperature) affects in the same way both thermal and charge conductivities.
The standard derivation of the WF law~\cite{AshcroftMermin,Mahan} ignores electron-electron (e-e) interactions, which can, in principle,  change the value of the WF ratio by affecting the charge and thermal conductivities in different ways.  Let us write these conductivities, at a finite frequency $\omega$,  in the common form
\begin{equation} \label{eq:general_Drude_conductivity}
\sigma_\ell (\omega) = \frac{ Q_\ell {\cal D}_{\ell}}{-i \omega + 1/\tau_{\ell}}\,,
\end{equation}
where  $\ell={\rm c}$ for the charge conductivity, $\ell= {\rm th}$ for the thermal conductivity, $Q_{\rm c}=e^2$ and $Q_{\rm th} = \pi k_{\rm B}^2 T/3$.  Here $\tau_{\rm c}$ and $\tau_{\rm th}$  are the  relaxation times  of charge and thermal currents respectively, and ${\cal D}_{\rm c}$ and ${\cal D}_{\rm th}$  are the  corresponding ``Drude weights".  Electron-electron interactions can modify the WF ratio by (i) creating a difference between the relaxation times $\tau_{\rm c}$ and $\tau_{\rm th}$  and/or (ii)  creating a difference between the Drude weights.  In general, the  ``amended" WF relation follows immediately from Eq.~(\ref{eq:general_Drude_conductivity}), and reads
\be\label{WFratio}
\frac{\sigma_{\rm th}}{\sigma_{\rm c} T} = \frac{\pi^2 k_{\rm B}^2}{3 e^2} 
\frac{{\cal D}_{\rm th}}{{\cal D}_{\rm c}} \frac{\tau_{\rm th}}{\tau_{\rm c}}
~.
\ee

Previous calculations of the WF ratio of a two-dimensional electron gas (2DEG) in the presence of e-e interactions~\cite{Castellani_prb_1986,Castellani_prl_1987,Arfi_jltp_1992,Schwab_annphys_2003,Raimondi_prb_2004,Catelani_JETP_2005} had focused on the diffusive regime, in which $\tau_{\rm th}$ and $\tau_{\rm c}$ are nearly identical and controlled by the electron-impurity scattering time $\tau$.  Therefore the renormalization factor ${\tau_{\rm th}}/{\tau_{\rm c}}$ is approximately $1$.  This, combined with the fact that the renormalization of the Drude weights appeared to be absent~\cite{Castellani_prb_1986,Castellani_prl_1987} (see, however, footnote~\cite{Drude_weight_renormalization} below), led to the conclusion that the WF law remains valid in the presence of e-e interactions.  

In this Letter we consider a different (``hydrodynamic") regime, which is expected to be relevant in very clean electronic systems, such as doped graphene,  with slowly varying potential modulations at not too low temperature.  In this regime, the e-e scattering time is much shorter than the electron impurity or electron-phonon scattering times~\cite{Andreev2011}.  Then a large difference can appear between the charge current  and the thermal current relaxation times, with the former being much larger than the latter.  This happens because e-e interactions  do not contribute to  charge current relaxation.  Indeed,  in each e-e scattering event the total momentum and hence the total charge current is conserved if umklapp processes are neglected.  Interband processes, potentially important in doped graphene, can be shown not to affect the charge conductivity in the Fermi liquid regime~\cite{Principi_charge_spin_conductivities}. In contrast to this,  the thermal current is not conserved in an e-e scattering process.  Therefore, contrary to the charge case, the thermal relaxation rate has a contribution from e-e interactions, {\it i.e.} $1/\tau_{\rm th} = 1/\tau + 1/\tau_{\rm th}^{\rm ee}$.  Below we prove that $1/\tau^{\rm ee}_{\rm th}$, at variance with $1/\tau_{\rm c}$, remains finite in the clean limit and equals $8/5$ of the quasiparticle decay rate.  Since the latter is always finite at finite temperature,  we conclude that the WF ratio, renormalized by $\tau_{\rm th}/\tau_{\rm c}\simeq (1 + \tau/\tau_{\rm th}^{\rm ee})^{-1}$, can become  arbitrarily small in the clean limit ($\tau \to \infty$). 

As for the second effect mentioned above, namely the different renormalization of the charge and thermal Drude weights, we will show that while the charge Drude weight is affected  by self-energy corrections~\cite{abedinpour_prb_2011} (reflected in a renormalized Fermi velocity) {\it and} by vertex corrections (described by the Landau Fermi liquid parameter $F_1^s$~\cite{Giuliani_and_Vignale,Pines_and_Nozieres}), the thermal Drude weight is affected only by self-energy corrections.  This leads to a further renormalization  ${\cal D}_{\rm th}/{\cal D}_{\rm c} \simeq (1+F_1^s)^{-1}$ of the WF ratio, which however is expected to be a small correction~\cite{Drude_weight_renormalization}.

We note that a hydrodynamic theory of the thermoelectric transport in graphene was worked out in Refs.~\cite{Muller_prb_2008_1,Muller_prb_2008_2,Foster_prb_2009}. There it was also shown that the thermal conductivity is finite in the presence of e-e interactions. However, no explicit form of the thermal transport time and Drude weight was given.  The present work is clearly within the framework of the Landau theory of Fermi liquids.  Mode-coupling effects~\cite{Pomeau_PhysRep_1975,Dorfman_AnnuRevPhysChem_1994,Kirkpatrick_JStatPhys_1997,Belitz_prb_1997,Belitz_prb_1998,Narayan_prl_2002,Belitz_rmp_2005,Meier_prb_2011,Belitz_prb_2012,Belitz_prb_2014} that could lead to a scale dependence of the electronic thermal conductivity -- such as the logarithmic dependence on system size observed for the thermal conductivity from phonons in pristine graphene~\cite{Xu_NatureComm_2014} -- are beyond the reach of such a theory.

\section{Model and  calculations}
Electrons and holes in doped graphene are described by the massless-Dirac-fermion (MDF) Hamiltonian (per spin and valley flavor -- hereafter $\hbar = 1$)~\cite{castroneto_rmp_2009,kotov_rmp_2012}
\begin{equation} \label{eq:MDF_Hamiltonian}
{\hat {\cal H}} = \sum_{{\bm k},\lambda} \varepsilon_{{\bm k},\lambda} {\hat \psi}^\dagger_{{\bm k},\lambda} {\hat \psi}_{{\bm k},\lambda}
+ \frac{1}{2} \sum_{{\bm q}} v_{\bm q} ({\hat n}_{{\bm q}} {\hat n}_{-{\bm q}} - {\hat n}_{{\bm 0}})
~,
\end{equation}
where $\psi_{{\bm k},\lambda}$ ($\psi^\dagger_{{\bm k},\lambda}$) destroys (creates) a particle with momentum ${\bm k}$ in band $\lambda=\pm$, and $\varepsilon_{{\bm k},\lambda} = \lambda v_{\rm F} k$ is the band energy.  $v_{\bm q} = (2\pi e^2)/(\epsilon q)$ is the non-relativistic Coulomb interaction,  $\epsilon$ is the dielectric constant of the environment, and ${\hat n}_{\bm q}$ is the density operator.  The Fermi energy is   $\varepsilon_{\rm F} = \pm \hbar v_{\rm F} k_{\rm F}$ ($+$ for electrons, $-$ for holes), where $k_{\rm F} = \sqrt{4\pi n/N_{\rm f}}$ is the Fermi wavevector, $n$ is the carrier density, and $N_{\rm f}=4$ is the number of spin-valley fermion flavors. Owing to the particle-hole symmetry of the model, we will consider exclusively $n$-type doping from now on.

%%%%%%%%%%%%%%%%%%%
\begin{figure}[t]
\begin{center}
\begin{tabular}{c}
\includegraphics[width=0.99\columnwidth]{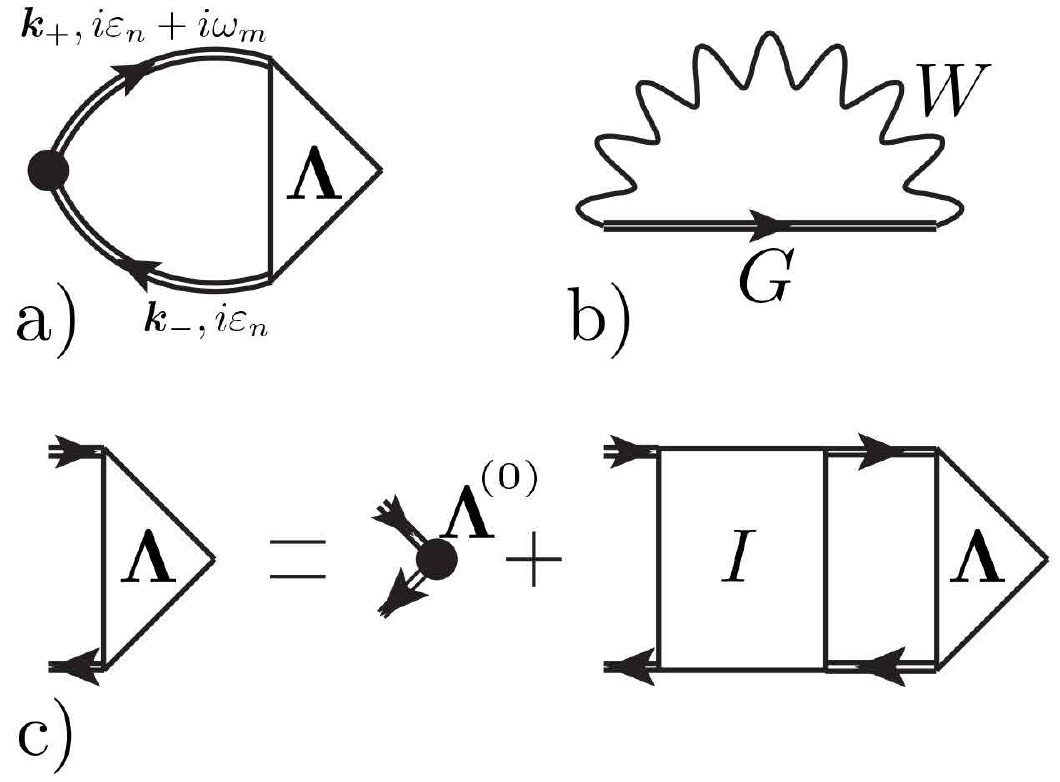}
\end{tabular}
\end{center}
\caption{
a) A diagrammatic picture of the calculation of the thermal-current-current response function. Panel a) shows the thermal-current linear response function. The solid dot on the left-hand side represents the bare current vertex $\Lambdabold^{(0)}$, and is connected to the dressed vertex $\Lambda$ by the two Green's functions depicted with double oriented lines. Panel b) shows the GW self-energy which renormalizes the Green's functions. The choice of the self energy uniquely determines, provided the Ward identities to be fulfilled~\cite{Giuliani_and_Vignale}, the self-consistent Bethe-Salpeter equation satisfied by the dressed vertex [panel c)], {\it i.e.} the irreducible interaction $I$.
\label{fig:one}}
\end{figure}
%%%%%%%%%%%%%%%%%%%

The thermal conductivity is defined in terms of the thermal-current linear response function $\chi_{J^{(Q)}_\alpha J^{(Q)}_\beta} ({\bm q}, \omega)$ as
\begin{eqnarray} \label{eq:thermal_c_def}
\sigma_{\rm th} = \lim_{\omega\to 0}\left[\frac{i}{\omega T} \chi_{J^{(Q)}_\alpha J^{(Q)}_\alpha} ({\bm q}={\bm 0}, \omega)\right]
~,
\end{eqnarray}
where $\alpha, \beta = x, y$, and the thermal current operator is
\be \label{ThermalCurrentOperator}
{\hat {\bm J}}^{(Q)}_{{\bm q}} = \sum_{{\bm k},\lambda,\lambda'} {\bm J}_{{\bm k}_-,{\bm k}_+}^{\lambda\lambda'}
\big(\partial_t{\hat \psi}^\dagger_{{\bm k}_-,\lambda} {\hat \psi}_{{\bm k}_+,\lambda'} + {\hat \psi}^\dagger_{{\bm k}_-,\lambda} \partial_t{\hat \psi}_{{\bm k}_+,\lambda'}\big)
~.
\ee
Here ${\bm J}_{{\bm k}_-,{\bm k}_+}^{\lambda\lambda'}$ is the matrix element of the number-current operator~\cite{SOM}, ${\bm k}_\pm = {\bm k} \pm {\bm q}/2$ and  $\partial_t \hat \psi_{{\bm k},\lambda}\equiv i[\hat{\cal H},\hat \psi_{{\bm k},\lambda}]$. 

Fig.~\ref{fig:one} summarizes the all-order diagrammatic re-summation needed to calculate the thermal conductivity. The thermal-current linear response function $\chi_{J^{(Q)}_\alpha J^{(Q)}_\beta} ({\bm q}, \omega)$ is given by the diagram depicted in Fig.~\ref{fig:one}a). Double solid lines represent Green's functions dressed by the ``GW" self-energy insertions of Fig.~\ref{fig:one}b). 
The bare current vertex [solid dot in Fig.~\ref{fig:one}a)] is~\cite{Castellani_prb_1986,Castellani_prl_1987,Arfi_jltp_1992,Schwab_annphys_2003,Raimondi_prb_2004,Catelani_JETP_2005}
\be\label{BareVertex}
\Lambdabold^{(0)}_{\lambda\lambda'}({\bm k}_-,i\varepsilon_n,{\bm k}_+,i\varepsilon_n+i\omega_m) =
(\varepsilon_n + \omega_m/2) {\bm J}^{\lambda\lambda'}_{{\bm k}_-,{\bm k}_+}~, 
\ee
where $\varepsilon_n$ and $\omega_m$ are, respectively, fermionic and bosonic Matsubara frequencies~\cite{Giuliani_and_Vignale}. The dressed current vertex ${\bm \Lambda}$, represented as a triangle in  Fig.~\ref{fig:one}, is determined by solving the self-consistent Bethe-Salpeter equation of Fig.~\ref{fig:one}c). 
The choice of the GW self-energy and the requirement of fulfilling the Ward identities uniquely determine the irreducible interaction $I$ of Fig.~\ref{fig:one}c)~\cite{Giuliani_and_Vignale,SOM}. 
After the analytical continuation to real frequencies~\cite{Bruus_and_Flensberg,Principi_charge_spin_conductivities}, the product of the two Green's functions appearing in $\chi_{J^{(Q)}_{\alpha} J^{(Q)}_{\beta}} ({\bm q}, \omega)$ is decomposed into products of advanced-advanced (schematically $G^A G^A$), retarded-retarded ($G^R G^R$) and advanced-retarded ($G^A G^R$) Green's function. 
$G^A G^A$ and $G^R G^R$ have poles on the same half of the complex plane, and give a vanishing contribution in the limit $(\varepsilon_{\rm F} \tau_{\rm qp}^{\rm ee})^{-1} \to 0$,~\cite{Bruus_and_Flensberg,Principi_charge_spin_conductivities} where $\tau_{\rm qp}^{\rm ee}$ is the lifetime of a quasiparticle at the Fermi surface.~\cite{SOM} 
In the limit $v_{\rm F} q \ll \omega, 1/\tau_{\rm qp}^{\rm ee} \ll \varepsilon_{\rm F}$ we approximate~\cite{Principi_charge_spin_conductivities} 
$
G^{({\rm A})}_\lambda G^{({\rm R})}_{\lambda'} \to
-2 i \delta_{\lambda=\lambda'=+} \Im m G^{({\rm R})}_{+}/(\omega + i/\tau_{\rm qp}^{\rm ee})
$.
In so doing  we neglect the incoherent part of  the Green's function, i.e., the part of $G$ that is not included in the quasiparticle-pole approximation.  Herein lies our Fermi liquid approximation.

The details of our calculation are given in the supplementary material, but we summarize the main conceptual steps here.
First, we remind the reader that at the non-interacting level 
\begin{eqnarray} \label{eq:thermal_c_0}
\sigma_{\rm th}^{(0)}(\omega) &=&
\frac{\pi^2 k_{\rm B}^2 T}{3} \frac{{\cal D}^{(0)}_{\rm th}}{-i \omega + \eta}
~,
\end{eqnarray}
where  $\eta=0^+$ and ${\cal D}^{(0)}_{\rm th} = N_{\rm f} \varepsilon_{\rm F}/(4\pi\hbar^2)$ is the non-interacting thermal Drude weight, which coincides with the non-interacting charge Drude weight ${\cal D}^{(0)}_{\rm c}$. $\sigma_{\rm th}^{(0)}(\omega)$ is  infinite in the limit $\omega \to 0$. 
The effect of e-e interactions is twofold. On the one hand, self-energy corrections replace $\eta \to 1/\tau_{\rm qp}^{\rm ee}$ on the right-hand side of Eq.~(\ref{eq:thermal_c_0}). On the other hand,  the vertex corrections multiplies the thermal conductivity by a factor $\gamma(\omega)$, defined as the ratio of the interacting vertex to the non interacting one:
$
\Lambdabold_{++} ({\bm k}, \varepsilon_++\omega,{\bm k},\varepsilon_-) = \gamma (\omega) \Lambdabold^{(0)}_{++} ({\bm k},\varepsilon_++\omega,{\bm k},\varepsilon_-) 
$, where all the wave vectors have fixed magnitude $\sim k_F$ and $\varepsilon_\pm = \varepsilon \pm i \eta$.  

The factor $\gamma(\omega)$ is determined from the solution of the Bethe-Salpeter equation~\cite{SOM}
\begin{eqnarray} \label{eq:T_Bethe_Salpeter}
&& \!\!\!\!\!\!\!\!
\Lambdabold_{++} ({\bm k}, \varepsilon_++\omega,{\bm k},\varepsilon_-) = \Lambdabold^{(0)}_{++} ({\bm k},\varepsilon_++\omega,{\bm k},\varepsilon_-)
\nonumber\\
&+&
\sum_{{\bm k}'} \int \frac{d\varepsilon'}{2\pi} I({\bm k},{\bm k'},\varepsilon,\varepsilon')
\Lambdabold_{++} ({\bm k}', \varepsilon'_++\omega, {\bm k}',\varepsilon'_-)
~,
\nonumber\\
\end{eqnarray}
in the limit $\omega, 1/\tau_{\rm qp}^{\rm ee} \ll \varepsilon_{\rm F}$ and to first order in $\varepsilon$. The result is
$
\gamma (\omega) = (\omega+i/\tau_{\rm qp}^{\rm ee})/(\omega + i/\tau_{\rm th}^{\rm ee})
$,
where the thermal transport time $\tau_{\rm th}^{\rm ee}$ is related to the quasiparticle lifetime by a simple proportionality factor:~\cite{SOM}
\begin{eqnarray}\label{eq:thermal_transport_time}
\tau_{\rm th}^{\rm ee} = \frac{5}{8} \tau_{\rm qp}^{\rm ee}\,.
\end{eqnarray}
Thus, our  result for the thermal conductivity is 
\begin{eqnarray} \label{eq:thermal_c}
\sigma_{\rm th}(\omega) &=&
\frac{\pi^2 k_{\rm B}^2 T}{3} \frac{{\cal D}^{(0)}_{\rm th}}{-i \omega + 8/(5 \tau_{\rm qp}^{\rm ee})}
~,
\end{eqnarray}
Notice that the Drude weight remains {\it unrenormalized} at this level of approximation.
At low temperature Eq.~(\ref{eq:thermal_c}) yields~\cite{Principi_charge_spin_conductivities,Li_prb_2013,Polini_QP_lifetime}
\begin{equation} \label{eq:invTau_asymptotics}
\frac{1}{\tau_{\rm qp}^{\rm ee}} \to \frac{4}{3} \frac{\pi}{N_{\rm f}} \frac{(k_{\rm B} T)^2}{\varepsilon_{\rm F}}
\ln \left(\zeta \frac{k_{\rm B} T}{\varepsilon_{\rm F}}\right)
~,
\end{equation}
with $\zeta= \pi/\sqrt{5}$.~\cite{SOM} As compared with Refs.~\cite{Principi_charge_spin_conductivities,Li_prb_2013,Polini_QP_lifetime}, Eq.~(\ref{eq:invTau_asymptotics}) shows an extra factor $4/3$ which is due to the definition of the quasiparticle decay rate adopted in this paper.~\cite{SOM}

\section{Thermal conductivity}
The key results~(\ref{eq:thermal_c}) and~(\ref{eq:invTau_asymptotics}) exhibit several remarkable features. 
The thermal relaxation rate, as the quasiparticle decay rate, is independent of the e-e coupling constant $\alpha_{\rm ee}$. This feature can be understood as follows. At low temperature the dominant contribution to $1/\tau_{\rm qp}^{\rm ee}$ is due to the collinear scattering of quasiparticles, whose phase space diverges~\cite{Muller_prl_2009} as a consequence of the linear dispersion of the MDF model. This seems to imply a divergence of $1/\tau_{\rm qp}^{\rm ee}$. However, the same phase-space divergence strongly enhances the screening of e-e interactions~\cite{Tomadin_prb_2013}. The strong screening in turn (i) translates the divergence of the quasiparticle decay rate into a weak (logarithmic) enhancement, and (ii) leads to an effective e-e interaction which is independent of $\alpha_{\rm ee}$.

%%%%%%%%%%%%%%%%%%%%
\begin{figure}[t]
\begin{center}
\begin{tabular}{c}
\includegraphics[width=0.99\columnwidth]{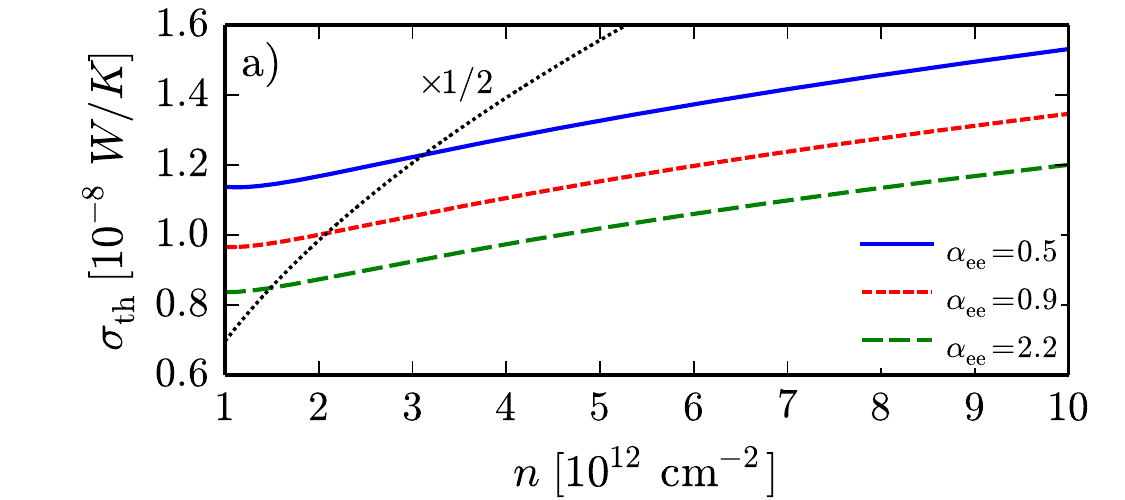}
\\
\includegraphics[width=0.99\columnwidth]{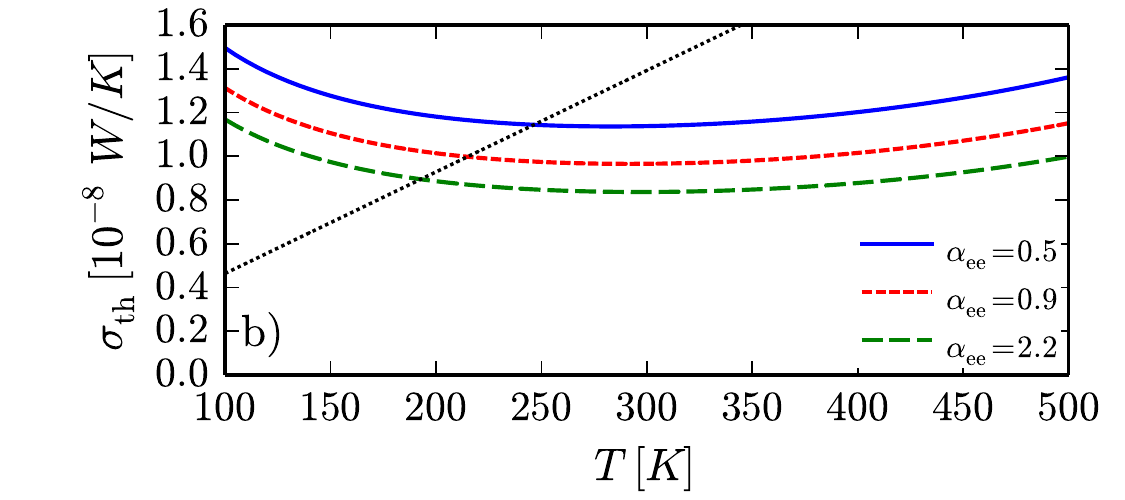}
\end{tabular}
\end{center}
\caption{
(Color online) Panel a) the electronic component of the thermal conductivity of MDFs $\sigma_{\rm th}$, as defined in Eq.~(\ref{eq:thermal_c}) and for $\omega=0$, plotted in units of $10^{-8}~{\rm W/K}$ as a function of the density $n$ (measured in units of $10^{12}~{\rm cm}^{-2}$). In this plot we fixed the temperature $T = 300~{\rm K}$. We show three curves for different values of the dimensionless coupling constant of e-e interactions, {\it i.e.} $\alpha_{\rm ee}=0.5$ (solid line), $\alpha_{\rm ee}=0.9$ (short-dashed line), and $\alpha_{\rm ee}=2.2$ (long-dashed line). Panel b) same as in panel a) but shown as a function of temperature (in units of ${\rm K}$) for a fixed excess carrier density $n=10^{12}~{\rm cm}^{-2}$ (corresponding to a Fermi temperature $T_{\rm F} \sim 1,300~{\rm K}$). As a comparison, in both panels we plot the thermal conductivity due to the impurity scattering (dotted line) of a MDF liquid whose mobility is $\mu=10,000~{\rm cm}^2/({\rm Vs})$. Note that in panel a) this curve is multiplied by a factor $1/2$. 
\label{fig:two}}
\end{figure}
%%%%%%%%%%%%%%%%%%%%

In Fig.~\ref{fig:two} we plot the d.c. thermal conductivity as a function of the carrier density at a fixed temperature $T=300~{\rm K}$ [panel a)], and as a function of temperature for a fixed carrier density $n=10^{12}~{\rm cm}^{-2}$ [panel b)], corresponding to a Fermi temperature $T_{\rm F} \sim 1,300~{\rm K}$. 
For $n \sim 10^{12}~{\rm cm}^{-2}$ and at room temperature, we find $\tau_{\rm th}^{\rm ee} \sim 0.1~{\rm ps}$, corresponding to a d.c. thermal conductivity of $~10^{-8}$ W/K. This value should be compared with the typical electron transport time $\tau$ due to impurities (or phonons). At the same carrier density and for a sample mobility $\mu = 10,000~{\rm cm}^2/({\rm V s})$, $\tau \sim 0.12~{\rm ps}$. Thus the WF ratio of Eq.~(\ref{WFratio}) is corrected by a factor $\tau_{\rm th}/\tau_{\rm c}\simeq (1 + \tau/\tau_{\rm th}^{\rm ee})^{-1} \sim 1/2$ with respect to its bare value. Typical samples should therefore show a clear violation of the Wiedemann-Franz law, provided one can separate out the phonon contribution to the thermal conductivity (see concluding remarks).

\section{The renormalization of the thermal Drude weight}
To calculate the renormalization of the Drude weight, we consider the kinetic equation for the distribution function of quasiparticles in the presence of a small temperature gradient that oscillates in time at a finite frequency $\omega$.  
The effective Hamiltonian for quasiparticles is~\cite{Pines_and_Nozieres,Conti_prb_1999}
$
{\hat {\cal H}}_{\rm qp}({\bm r},{\bm k}, t) = \xi_{\bm k}^\star + \sum_{{\bm k}'} f_{{\bm k},{\bm k}'} n_{1,{\bm k}'}({\bm r},t)
$, where $\xi_{{\bm k}}^\star = v_{\rm F}^\star k - \mu^\star$ is the renormalized quasiparticle energy measured from the renormalized chemical potential $\mu^\star$,  $f_{{\bm k},{\bm k}'}$ is the Landau interaction function~\cite{Giuliani_and_Vignale}, and $n_{1,{\bm k}}({\bm r},t)$ represents the departure of the quasiparticle distribution function $n_{{\bm k}}({\bm r},t)$ from the equilibrium one $n_0(\xi_{\bm k}^\star)$.  
Neglecting the collision integral, which plays no role in the calculation of the Drude weight, we have
\begin{eqnarray} \label{eq:EOM_qp}
&& \!\!\!\!\!\!\!\!
({\bm q}\cdot {\bm v}_{\bm k}^\star - \omega) n_{1,{\bm k}} - n_0'(\xi_{\bm k}^\star) {\bm v}_{\bm k}^\star \cdot\Bigg[ {\bm q} \sum_{{\bm k}'} f_{{\bm k},{\bm k}'} n_{1,{\bm k}'} + \xi_{\bm k}^\star \frac{{\bm \nabla}T}{T}\Bigg] 
\nonumber\\
&=& 0\,,
\end{eqnarray}
where $n_{1,{\bm k}} \equiv n_{1,{\bm k}}({\bm q},\omega)$ is the Fourier transform of the distribution function at the wave vector ${\bf q}$ of the disturbance, and $v_{\bm k}^\star = {\bm \nabla}_{\bm k} \xi_{\bm k}^\star$ is the quasiparticle velocity.  $n_0'(\xi_{\bm k}^\star) $ is the derivative of the Fermi distribution function with respect to $\xi_{\bm k}^\star$.  To ${\cal O}(q^2/\omega^2)$, Eq.~(\ref{eq:EOM_qp}) is solved by the {\it Ansatz} $n_{1,{\bm k}}({\bm q},\omega) =  n_0'(\xi_{\bm k}^\star) A_{\bm k}({\bm q},\omega) \xi_{\bm k}^\star$, where $A_{\bm k}({\bm q},\omega)$ is to be determined. When ${\bm k}$ and ${\bm k}'$ are both at the Fermi surface, we can assume that~\cite{Giuliani_and_Vignale} $f_{{\bm k},{\bm k}'}$ is a function only of the angle between ${\bm k}$ and ${\bm k}'$.  Inserting the trial function $n_{1,{\bm k}}({\bm q},\omega)$  we see that the first term in the square brackets of Eq.~(\ref{eq:EOM_qp}) vanishes at order $T^2$ due to the strong cancellation of  contributions from opposite sides of the Fermi surface (the  $\xi_{\bm k}^\star$  factor is antisymmetric). This is the cancellation of vertex corrections we anticipated in the introduction.  Essentially the same cancellation occurs in the classic calculation of the heat capacity in the Landau Fermi liquid theory~\cite{Giuliani_and_Vignale,Pines_and_Nozieres}. The absence of vertex corrections is  confirmed by the calculation of the thermal Drude weight to first order in the strength of e-e interactions performed in~\cite{SOM}. We are thus left with
\begin{equation}
n_{1,{\bm k}} = -n_0'(\xi_{\bm k}^\star)\frac{({\bm q}\cdot{\bm v}_{\bm k}^\star)(v_{\bm k}^\star\cdot{\bm \nabla}T)}{\omega^2 T} \xi_{\bm k}^\star
~.
\end{equation}
The induced variation of the entropy of quasiparticles is given by~\cite{entropy_density} $\delta S({\bm q},\omega) = \sum_{\bm k} n_{1,{\bm k}} \xi_{\bm k}^\star/(k_{\rm B} T)$. From this we readily extract the thermal Drude weight via the relation $\delta S({\bm q},\omega) =(\pi^2 k_B^2T/3){\cal D}_{\rm th}q^2/\omega^2$.  The result is
\be \label{eq:FL_th_Drude_weight}
{\cal D}_{\rm th} =  \frac{N_{\rm f}}{4\pi \hbar} k_{\rm F} v_{\rm F}^\star
~,
\ee
where many-body effects enter only through the renormalization of the Fermi velocity. This should be contrasted with the charge Drude weight, for which the vertex correction does not vanish, yielding~\cite{abedinpour_prb_2011,Levitov_prb_2013} 
\be
{\cal D}_{\rm c} = \frac{N_{\rm f}}{4\pi \hbar} k_{\rm F} v_{\rm F}^\star (1+F_1^{\rm s})
~,
\ee
where $F_1^{\rm s}$ is the first spin-symmetric Landau parameter~\cite{Giuliani_and_Vignale}. We conclude that the WF ratio, Eq.~(\ref{WFratio}) is further renormalized by a factor  ${\cal D}_{\rm th}/{\cal D}_{\rm c} = (1+F_1^{\rm s})^{-1}$. In a two-dimensional electron gas $1+F^{\rm s}_1$ is always very close to one in a broad range of values of the strength of e-e interactions~\cite{Giuliani_and_Vignale}. We expect this to be true also in graphene in the Fermi-liquid regime. This leaves  the renormalization $\tau_{\rm th}/\tau_{\rm c}$ as the main factor controlling the value of the WF ratio, which can be made arbitrarily small by increasing the sample purity.

\section{Summary and conclusions}
According to the Wiedemann-Franz law, the charge ($\sigma_{\rm c}$) and thermal ($\sigma_{\rm th}$) conductivities of a Fermi liquid satisfy the relation $\sigma_{\rm th} = L T \sigma_{\rm c}$, where the Lorenz number $L = \pi^2 k_{\rm B}^2/(3 e^2)$. The charge conductivity of doped graphene is calculated in the limit of $T\to 0$ in Ref.~\cite{Principi_charge_spin_conductivities}, where it is found to diverge faster than $1/T^2$ in absence of impurities and phonons. At low temperature doped graphene is an effectively Galilean invariant system, and a homogeneous current cannot relax efficiently.  In contrast to this, the thermal conductivity -- Eq.~(\ref{eq:thermal_c}) -- is  always finite as long as $T \neq 0$, and diverges only in the zero-temperature limit as $\sigma_{\rm th} \propto T^{-1}/\ln(T)$. This result manifestly violates the WF law. In the presence of electron-electron interactions, the WF ratio of Eq.~(\ref{WFratio}) is renormalized by a factor $\tau_{\rm th}/\tau_{\rm c}\simeq (1 + \tau/\tau_{\rm th}^{\rm ee})^{-1}$ with respect to its bare value. For an electron density $n\sim 10^{12}~{\rm cm}^{-2}$, a mobility $\mu \sim 10,000~{\rm cm}^{2}/({\rm Vs})$, and $T=300~{\rm K}$, $\tau_{\rm th}/\tau_{\rm c}\sim 1/2$. 

Experimentally, it is quite challenging to separate the electronic contribution from the phononic (pristine) contribution to the thermal conductivity. The latter is expected to be larger than the former. Indeed, for $n\sim 10^{12}~{\rm cm}^{-2}$ and $T=300~{\rm K}$ the electronic contribution to the thermal conductivity is $\sigma_{\rm th}\sim 10^{-8}~{\rm W/K}$. Under the same conditions, the phononic contribution is $\sigma_{\rm th}^{({\rm ph})} \sim 10^{-7}~{\rm W/K}$, and can reach values as high as $\sigma_{\rm th}^{({\rm ph})} \sim 10^{-6}~{\rm W/K}$ in suspended samples~\cite{Balandin_nanolett_2008,Seol_science_2010,Chen_ACSnano_2011,Balandin_natmat_2011,Pop_MRS_2012,Xu_NatureComm_2014}. However, this contribution vanishes as the temperature is lowered~\cite{Xu_NatureComm_2014}.
Thus, the separation can be achieved by tracking the dependence of the thermal conductivity on the carrier density and temperature, and/or by encasing graphene~\cite{Jang_nanolett_2010} or placing it on substrates like ${\rm SiO}_2$~\cite{Liao_prl_2011} or h-BN which suppress the large contribution of flexural phonons to the thermal conductivity~\cite{Mariani_prl_2008,Lindsay_prb_2010}.

\section{Acknowledgments}
This work was supported in part by DOE grant DE-FG02-05ER46203 and by a Research Board Grant at the University of Missouri.

\appendix

\section{Introduction}
\begin{figure}[t]
\begin{center}
\begin{tabular}{c}
\includegraphics[width=0.99\columnwidth]{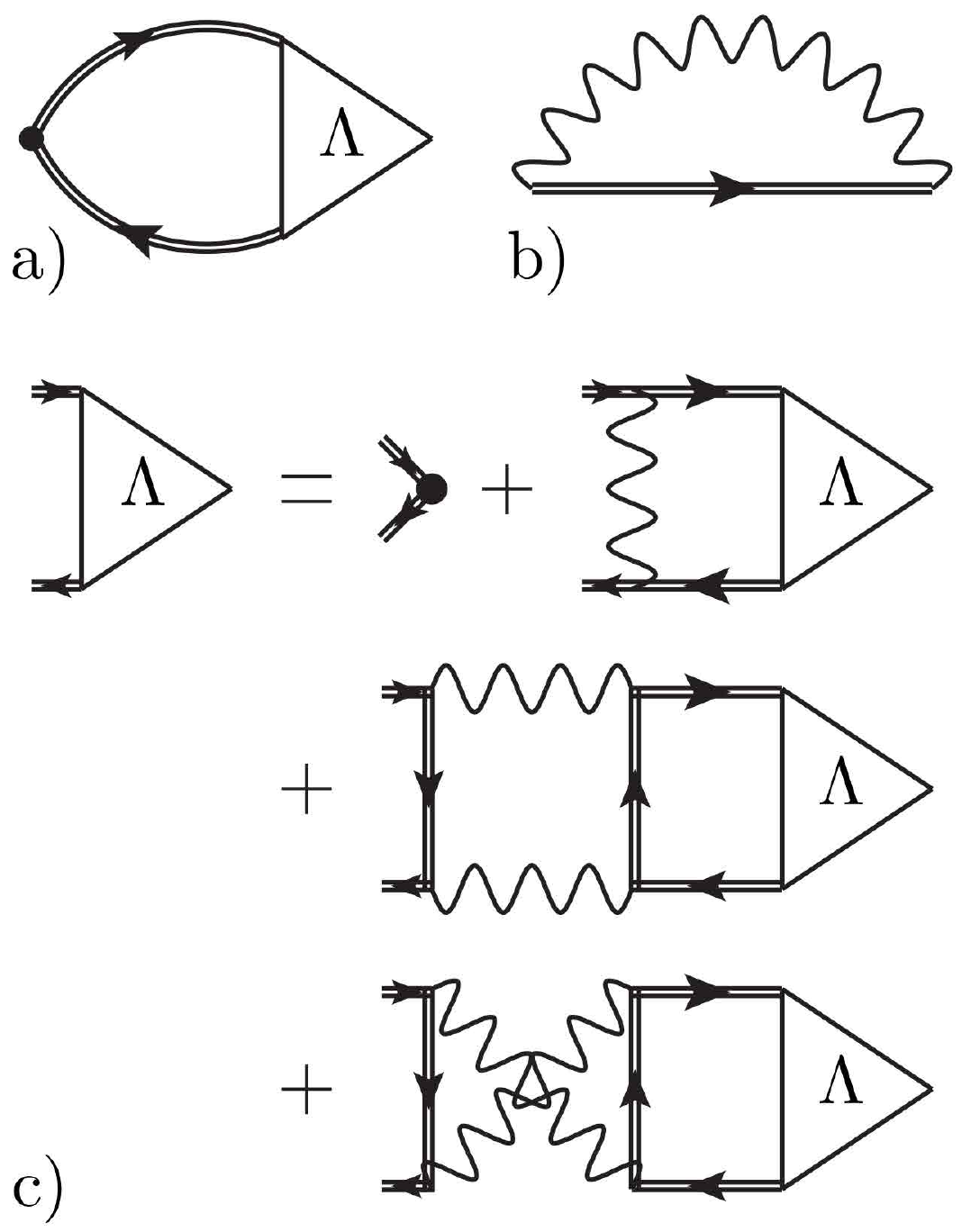}
\end{tabular}
\end{center}
\caption{
a) A diagrammatic picture of the calculation of the thermal-current-current response function. Panel a) shows the thermal-current linear response function. The solid dot on the left-hand side represents the bare current vertex ${\bm \Lambda}^{(0)}$, and is connected to the dressed vertex ${\bm \Lambda}$ by the two Green's functions depicted with double oriented lines. Panel b) shows the GW self-energy which renormalizes the Green's functions. The choice of the self energy uniquely determines, provided the Ward identities to be fulfilled~\cite{Giuliani_and_Vignale}, the self-consistent Bethe-Salpeter equation satisfied by the dressed vertex [panel c)].
\label{fig:SM_one}}
\end{figure}

The calculation of the thermal conductivity due to e-e interactions requires the knowledge of the self-energy [Fig.~\ref{fig:SM_one}b)] and vertex [Fig.~\ref{fig:SM_one}c)] corrections to the energy-current response function [Fig.~\ref{fig:SM_one}a)]. In what follows we start from the evaluation of the self-energy corrections, which are encoded in the finite quasiparticle lifetime at the Fermi surface. We then proceed to the calculation of the vertex correction.

%%%%%%%%%%%%%%%%%%%
%%%%%%%%%%%%%%%%%%%

\section{Self-energy correction --- the quasiparticle lifetime at the Fermi surface} \label{app:QP_lifetime}
In this section we derive the quasiparticle lifetime at the Fermi surface. Let us recall the expression of the GW self-energy [Fig.~\ref{fig:SM_one}b)] which reads
\begin{eqnarray} \label{eq:SM_app_GW}
\Sigma_\lambda({\bm k},i\varepsilon_n) &=& -k_{\rm B} T \sum_{{\bm k}',\lambda'} \sum_{\varepsilon_{n'}} W({\bm k}'-{\bm k},i\varepsilon_{n'} - i\varepsilon_n)
\nonumber\\
&\times&
G_{\lambda'}({\bm k}',i\varepsilon_{n'}) {\cal D}_{\lambda\lambda'}({\bm k},{\bm k}') {\cal D}_{\lambda'\lambda}({\bm k}',{\bm k})
~,
\nonumber\\
\end{eqnarray}
where
\begin{equation} \label{eq:SM_app_W_RPA}
W({\bm q},i\omega_{m}) = \frac{v_{\bm q}}{1-v_{\bm q} \chi_{nn}({\bm q},i\omega_m)}
\end{equation}
is the dynamically screened Coulomb interaction.
Here $v_{\bm q} = 2\pi e^2/(\epsilon q)$ is the bare Coulomb interaction, $\epsilon$ is the dielectric constant of the environment, and $\chi_{nn}^{(0)}({\bm q},i\omega_m)$ is the non-interacting density-density response function, which is
\begin{eqnarray} \label{eq:SM_app_chi_nn}
\chi_{nn}({\bm q},i\omega_m) &=& N_{\rm f} k_{\rm B} T \sum_{{\bm q}',\varepsilon_{n''}} \sum_{\lambda'',\mu''} G_{\lambda''}({\bm q}',i\varepsilon_{n''})
\nonumber\\
&\times&
G_{\mu''}({\bm q'}+{\bm q}, i\varepsilon_{n''}+i\omega_m)
\nonumber\\
&\times&
{\cal D}_{\lambda''\mu''}({\bm q}',{\bm q}'+{\bm q}) {\cal D}_{\mu''\lambda''}({\bm q}'+{\bm q},{\bm q}') 
~.
\nonumber\\
\end{eqnarray}
Here $\varepsilon_n = (2n+1)\pi/\beta$ ($\omega_m = 2m\pi/\beta$) is a Fermionic (Bosonic) Matsubara frequency [$\beta = (k_{\rm B} T)^{-1}$], and
\begin{equation} \label{eq:SM_D_element}
{\cal D}_{\lambda\lambda'} ({\bm k}, {\bm k}') =
\frac{e^{i(\varphi_{\bm k}-\varphi_{{\bm k}'})/2} + \lambda\lambda' e^{-i(\varphi_{\bm k}-\varphi_{{\bm k}'})/2}}{2}
%~,
\end{equation}
is the matrix element of the density operator between the eigenstates of the MDF Hamiltonian. In Eq.~(\ref{eq:SM_D_element}) $\varphi_{\bm k}$ is the angle between the momentum ${\bm k}$ and the ${\hat {\bm x}}$-axis. For future purposes we also define $\alpha_{\rm ee} = e^2/(\epsilon \hbar v_{\rm F})$ as the coupling constant of e-e interactions, and the matrix elements of the current operator
\begin{eqnarray} \label{eq:SM_J_element}
J^{\lambda\lambda'}_{{\bm k},{\bm k}',x} &=&
v_{\rm F} \frac{\lambda' e^{i(\varphi_{\bm k}+\varphi_{{\bm k}'})/2} + \lambda e^{-i(\varphi_{\bm k}+\varphi_{{\bm k}'})/2}}{2}
~,
\nonumber\\
J^{\lambda\lambda'}_{{\bm k},{\bm k}',y} &=& 
v_{\rm F} \frac{\lambda' e^{i(\varphi_{\bm k}+\varphi_{{\bm k}'})/2} - \lambda e^{-i(\varphi_{\bm k}+\varphi_{{\bm k}'})/2}}{2i}
~.
\end{eqnarray}

We first consider Eq.~(\ref{eq:SM_app_GW}), and we analytically continue it to real frequencies. We thus define $f_\Sigma(i\varepsilon_{n'} - i\varepsilon_n, i\varepsilon_{n'})$ such that
\begin{eqnarray}
\Sigma_\lambda({\bm k},i\varepsilon_n) &\equiv& -k_{\rm B} T \sum_{\varepsilon_{n'}} f_\Sigma(i\varepsilon_{n'} - i\varepsilon_n, i\varepsilon_{n'})
\nonumber\\
&=&
\oint_{\cal C} \frac{dz}{2\pi i} n_{\rm F}(z) f_\Sigma(z - i\varepsilon_n, z)
~.
\end{eqnarray}
The contour ${\cal C}$ in the complex plane encircles all the poles of the Fermi-Dirac distribution $n_{\rm F}(z) = \big[\exp(\beta z)+1\big]^{-1}$, and leaves outside the branch cuts of $f_\Sigma(z - i\varepsilon_n, z)$, which are parallel to the real axis and pass by $z=0, i \varepsilon_n$. Taking the limit $i\varepsilon_n \to \varepsilon + i\eta$ we obtain the retarded self-energy~\cite{Giuliani_and_Vignale}
\begin{eqnarray} \label{eq:SM_app_sigma_retarded}
\Sigma_\lambda({\bm k},\varepsilon_+) &=&
\int_{-\infty}^{\infty} \frac{d\varepsilon'}{2\pi i} \Big\{ 
\big[n_{\rm F}(\varepsilon') + n_{\rm B}(\varepsilon'-\varepsilon)\big] 
\nonumber\\
&\times&
\big[f_\Sigma(\varepsilon'_- - \varepsilon, \varepsilon'_+) - f_\Sigma(\varepsilon'_+ - \varepsilon, \varepsilon'_+)\big]
\nonumber\\
&+&
n_{\rm F}(\varepsilon') \big[ f_\Sigma(\varepsilon'_- - \varepsilon, \varepsilon'_-) - f_\Sigma(\varepsilon'_+ - \varepsilon, \varepsilon'_+) \big]
\Big\}
~.
\nonumber\\
\end{eqnarray}
Here $\varepsilon_\pm = \varepsilon \pm i \eta$ and we defined the Bose-Einstein distribution $n_{\rm B}(\varepsilon) = \big[\exp(\beta \varepsilon) - 1\big]^{-1}$. After the analytical continuation, $G_{\lambda}({\bm k},\varepsilon_+) = G^{({\rm R})}_{\lambda}({\bm k},\varepsilon)$ and $G_{\lambda}({\bm k},\varepsilon_-) = G^{({\rm A})}_{\lambda}({\bm k},\varepsilon)$. Here $G^{({\rm A})}_{\lambda}({\bm k}, \varepsilon)$ [$G^{({\rm R})}_{\lambda}({\bm k}, \varepsilon)$] represents the advanced [retarded] Green's function. 

Note that the term in the last line of Eq.~(\ref{eq:SM_app_sigma_retarded}) is purely imaginary and gives a purely real contribution to the self-energy after multiplication by the imaginary unit. Being interested in the imaginary part of the self-energy, we can neglect this term. One would be tempted, at this point, to approximate the imaginary part of the self-energy with its value at the Fermi surface. Such approximation is  too crude. A slightly more sophisticated approximation consists in writing
\begin{eqnarray} \label{eq:SM_app_taum1}
\frac{1}{\tau_{\rm qp}^{\rm ee}} &=& 2 \int_{-\infty}^{\infty} d\varepsilon \frac{\partial n_{\rm F}(\varepsilon)}{\partial \varepsilon} \Im m \big[\Sigma_+({\bm k},\varepsilon^+)\big]\Big|_{k=k_{\rm F}}
\nonumber\\
&=& - 2 \int_{-\infty}^{\infty} d\varepsilon \int_{-\infty}^{\infty} \frac{d\varepsilon'}{\pi} \frac{\partial n_{\rm F}(\varepsilon)}{\partial \varepsilon} \big[n_{\rm F}(\varepsilon') + n_{\rm B}(\varepsilon'-\varepsilon)\big] 
\nonumber\\
&\times&
\sum_{{\bm k}',\lambda'}\Im m W({\bm k}-{\bm k}', \varepsilon'_+ - \varepsilon) 
{\cal D}_{+\lambda'}({\bm k},{\bm k}') {\cal D}_{\lambda'+}({\bm k}',{\bm k})
~.
\nonumber\\
\end{eqnarray}
In this equation $|{\bm k}| = k_{\rm F}$ is understood. Moreover, Eq.~(\ref{eq:SM_app_taum1}) assumes that the energy-dependence of the Density-of-States (DOS) is negligible. This is not true in graphene, whose DOS scales linearly with energy. However, such density dependence is responsible for corrections to Eq.~(\ref{eq:SM_app_taum1}) which can be ignored in the low-temperature limit. The imaginary part of the screened e-e interaction is
\begin{equation} \label{eq:SM_app_Im_W}
\Im m W({\bm q}, \omega_+) = |W({\bm q}, \omega_+)|^2 \Im m \chi_{nn}({\bm q}, \omega_+)
~.
\end{equation}
To determine $\Im m \chi_{nn}({\bm q}, \omega_+)$, we first analytically continue $\chi_{nn}({\bm q}, i\omega_m)$ defined in Eq.~(\ref{eq:SM_app_chi_nn}) to real frequencies. In analogy to the self-energy we define
\begin{eqnarray} 
\chi_{nn}({\bm q},i\omega_m) &=& k_{\rm B} T \sum_{\varepsilon_{n''}} f_\chi(i\varepsilon_{n''} + i\omega_m, i\varepsilon_{n''})
\nonumber\\
&=&
-\oint_{{\cal C}'} \frac{dz}{2\pi i} n_{\rm F}(z) f_\chi(z + i\omega_m, z)
~.
\end{eqnarray}
The contour ${\cal C}'$ encircles only the poles of the Fermi function and excludes the branch cuts of $f_\chi(z + i\omega_n, z)$, which are parallel to the real axis and pass by $z=0, -i \omega_m$. 
Taking the limit $i\omega_m \to \omega+i\eta$ we get
\begin{eqnarray} 
\chi_{nn}({\bm q},\omega_+) &=&  -\int_{-\infty}^{\infty} \frac{d\varepsilon''}{2\pi i} \Big\{ \big[ n_{\rm F}(\varepsilon''+\omega) - n_{\rm F}(\varepsilon'') \big]
\nonumber\\
&\times&
\big[ f(\varepsilon''_++\omega,\varepsilon''_-) - f(\varepsilon''_-+\omega,\varepsilon''_-) \big]
\nonumber\\
&+&
n_{\rm F}(\varepsilon'') \big[ f(\varepsilon''_++\omega,\varepsilon''_+) - f(\varepsilon''_-+\omega,\varepsilon''_-) \big]
\Big\}
~.
\nonumber\\
\end{eqnarray}
Again the last term gives no contribution to the imaginary part of $\chi_{nn}({\bm q},\omega_+)$, which reads
\begin{eqnarray} \label{eq:SM_app_Im_chi_nn}
&& \!\!\!\!\!\!\!\!
\Im m\chi_{nn}({\bm q},\omega_+) = -N_{\rm f} \sum_{ {\bm q}', \lambda'',\mu''} \int \frac{d\varepsilon''}{\pi} \big[ n_{\rm F}(\varepsilon'') - n_{\rm F}(\varepsilon''+\omega) \big] 
\nonumber\\
&\times&
\Im m G_{\lambda''}({\bm q}',\varepsilon''_+)
\Im m G_{\mu''}({\bm q'}+{\bm q}, \varepsilon''_++\omega)
\nonumber\\
&\times&
{\cal D}_{\lambda''\mu''}({\bm q}',{\bm q}'+{\bm q}) {\cal D}_{\mu''\lambda''}({\bm q}'+{\bm q},{\bm q}') 
~.
\nonumber\\
\end{eqnarray}
We put Eq.~(\ref{eq:SM_app_Im_chi_nn}) into Eq.~(\ref{eq:SM_app_Im_W}) and~(\ref{eq:SM_app_taum1}), and we get
\begin{eqnarray} \label{eq:SM_app_taum1_2}
&& \!\!\!\!\!\!\!\!
\frac{1}{\tau_{\rm qp}^{\rm ee}} = 2N_{\rm f} \sum_{{\bm k}',{\bm q}'}\sum_{\lambda', \lambda'',\mu''} 
\int_{-\infty}^{\infty} d\varepsilon \int_{-\infty}^{\infty} \frac{d\varepsilon'}{\pi} \int_{-\infty}^{\infty} \frac{d\varepsilon''}{\pi}
\frac{\partial n_{\rm F}(\varepsilon)}{\partial \varepsilon} 
\nonumber\\
&\times&
\big[n_{\rm F}(\varepsilon') + n_{\rm B}(\varepsilon'-\varepsilon)\big] 
\big[ n_{\rm F}(\varepsilon''+\varepsilon) - n_{\rm F}(\varepsilon''+\varepsilon') \big] 
\nonumber\\
&\times&
|W({\bm k}-{\bm k}',\varepsilon'_+)|^2
\Im m G_+({\bm k}',\varepsilon'_+)
\nonumber\\
&\times&
\Im m G_{\lambda''}({\bm q}'-{\bm k},\varepsilon''_++\varepsilon)
\Im m G_{\mu''}({\bm q'}-{\bm k}', \varepsilon''_++\varepsilon')
\nonumber\\
&\times&
{\cal D}_{+\lambda'}({\bm k},{\bm k}') {\cal D}_{\lambda'+}({\bm k}',{\bm k})
{\cal D}_{\lambda''\mu''}({\bm q}'-{\bm k},{\bm q}'-{\bm k}')
\nonumber\\
&\times&
{\cal D}_{\mu''\lambda''}({\bm q}'-{\bm k}',{\bm q}'-{\bm k}) 
~.
\end{eqnarray}
We now observe that
\begin{eqnarray} \label{eq:SM_nB_nF_approx}
{\cal N} &\equiv&
\frac{\partial n_{\rm F}(\varepsilon)}{\partial \varepsilon} \big[n_{\rm F}(\varepsilon') + n_{\rm B}(\varepsilon'-\varepsilon)\big] 
\nonumber\\
&\times&
\big[ n_{\rm F}(\varepsilon''+\varepsilon) - n_{\rm F}(\varepsilon''+\varepsilon') \big] 
\nonumber\\
&=&
\frac{\partial n_{\rm B}(\varepsilon'')}{\partial \varepsilon''} \big[ n_{\rm F}(\varepsilon+\varepsilon'') - n_{\rm F}(\varepsilon) \big] 
\nonumber\\
&\times&
\big[ n_{\rm F}(\varepsilon'+\varepsilon'') - n_{\rm F}(\varepsilon') \big]
\nonumber\\
&\simeq&
\varepsilon''^2 \frac{\partial n_{\rm B}(\varepsilon'')}{\partial \varepsilon''} \frac{\partial n_{\rm F}(\varepsilon)}{\partial \varepsilon} \frac{\partial n_{\rm F}(\varepsilon')}{\partial \varepsilon'}\,,
\end{eqnarray}
where in the last line we expanded for small $\varepsilon''$.  Eq.~(\ref{eq:SM_app_taum1_2}) thus becomes
\begin{eqnarray} \label{eq:SM_app_taum1_3}
&& \!\!\!\!\!\!\!\!
\frac{1}{\tau_{\rm qp}^{\rm ee}} \to 2N_{\rm f} \sum_{{\bm k}',{\bm q}'}\sum_{\lambda', \lambda'',\mu''} 
\int_{-\infty}^{\infty} d\varepsilon \int_{-\infty}^{\infty} \frac{d\varepsilon'}{\pi} \int_{-\infty}^{\infty} \frac{d\varepsilon''}{\pi}
\frac{\partial n_{\rm B}(\varepsilon'')}{\partial \varepsilon''} 
\nonumber\\
&\times&
\varepsilon''^2 \frac{\partial n_{\rm F}(\varepsilon)}{\partial \varepsilon} \frac{\partial n_{\rm F}(\varepsilon')}{\partial \varepsilon'} 
|W({\bm k}-{\bm k}',\varepsilon'_+)|^2
\Im m G_+({\bm k}',\varepsilon'_+)
\nonumber\\
&\times&
\Im m G_{\lambda''}({\bm q}'-{\bm k},\varepsilon''_++\varepsilon)
\Im m G_{\mu''}({\bm q'}-{\bm k}', \varepsilon''_++\varepsilon')
\nonumber\\
&\times&
{\cal D}_{+\lambda'}({\bm k},{\bm k}') {\cal D}_{\lambda'+}({\bm k}',{\bm k})
{\cal D}_{\lambda''\mu''}({\bm q}'-{\bm k},{\bm q}'-{\bm k}')
\nonumber\\
&\times&
{\cal D}_{\mu''\lambda''}({\bm q}'-{\bm k}',{\bm q}'-{\bm k}) 
~.
\end{eqnarray}
In evaluating an integral of the form 
\begin{equation}
{\cal I} = \int_{-\infty}^{\infty} d\varepsilon'' \frac{\partial n_{\rm B}(\varepsilon'')}{\partial \varepsilon''} \varepsilon''^2 f(\varepsilon'')
~,
\end{equation}
where $f(\varepsilon'')$ is some smooth function of its argument, we exploit the fact that the weighting function $\varepsilon''^2\partial n_{\rm B}(\varepsilon'')/\partial \varepsilon''$  is strongly peaked at $\varepsilon''=0$ and its width scales with $k_{\rm B}^2 T^2/\varepsilon_{\rm F}$.  This does not mean, however, that one can simply replace $f(\varepsilon'')$ by $f(0)$.  Such a crude approximation of Eq.~(\ref{eq:SM_app_taum1_3}) would introduce a spurious divergence in the quasiparticle decay rate, because it spoils the subtle cancellation between two infinities which occur (i) in the phase space of the collinear scattering~\cite{Muller_prl_2009,Tomadin_prb_2013} and (ii) in the screening of e-e interactions. Both divergences are connected to the linear-in-momentum energy dispersion of massless Dirac fermions. The cancellation occurs as long as the argument of the function $f(\varepsilon)$ is finite.   To take this into account we approximate
\begin{equation} \label{eq:SM_energy_approx}
\int_{-\infty}^{\infty} d\varepsilon'' \frac{\partial n_{\rm B}(\varepsilon'')}{\partial \varepsilon''} \varepsilon''^2 f(\varepsilon'') = -\frac{2 \pi^2 (k_{\rm B} T)^2}{3} f({\bar \varepsilon}) + {\cal O}(T^4)\,,
\end{equation}
where ${\bar \varepsilon}$ can be estimated as
\begin{equation} \label{eq:SM_epsilon_bar}
{\bar \varepsilon} = \frac{1}{2} \sqrt{-\frac{3}{2\pi^2 (k_{\rm B} T)^2} \int_{-\infty}^{+\infty} d\varepsilon~ \varepsilon^4  \frac{\partial n_{\rm B}(\varepsilon)}{\partial \varepsilon} } = {\bar T} \varepsilon_{\rm F}\,.
\end{equation}
Here we have defined ${\bar T} = \zeta k_{\rm B} T/\varepsilon_{\rm F}$ and $\zeta= \pi/\sqrt{5}$.  The factor $-3/[2\pi^2 (k_{\rm B} T)^2]$ normalizes the weight of the function $\varepsilon^2\partial n_{\rm B}(\varepsilon)/\partial \varepsilon$ to one. We have thus taken ${\bar \varepsilon}$ to be half of the variance of the distribution $\varepsilon^2 \partial n_{\rm B}(\varepsilon)/\partial \varepsilon$.
With this approximation we finally get the quasiparticle lifetime at the Fermi surface:
\begin{eqnarray} \label{eq:SM_inverse_lifetime}
\frac{1}{\tau_{\rm qp}^{\rm ee}} 
&=&
- \frac{4}{3} N_{\rm f} (k_{\rm B} T)^2 \sum_{{\bm k}',{\bm q}'}
|W({\bm k}'-{\bm k},{\bar \varepsilon})|^2
\nonumber\\
&\times&
\Im m \big[G^{({\rm R},\sigma)}_{+}({\bm k}',{\bar \varepsilon})\big]
\Im m \big[G^{({\rm R},\sigma')}_{+}({\bm q}'-{\bm k},0)\big]
\nonumber\\
&\times&
\Im m \big[G^{({\rm R},\sigma')}_{+}({\bm q'}-{\bm k}', {\bar \varepsilon})\big]
{\cal D}_{++}({\bm k},{\bm k}') {\cal D}_{++}({\bm k}',{\bm k})
\nonumber\\
&\times&
{\cal D}_{++}({\bm q}'-{\bm k},{\bm q}'-{\bm k}') {\cal D}_{++}({\bm q}'-{\bm k}',{\bm q}'-{\bm k}) 
~.
\nonumber\\
\end{eqnarray}

\begin{figure}[t]
\begin{center}
\begin{tabular}{c}
\includegraphics[width=0.6\columnwidth]{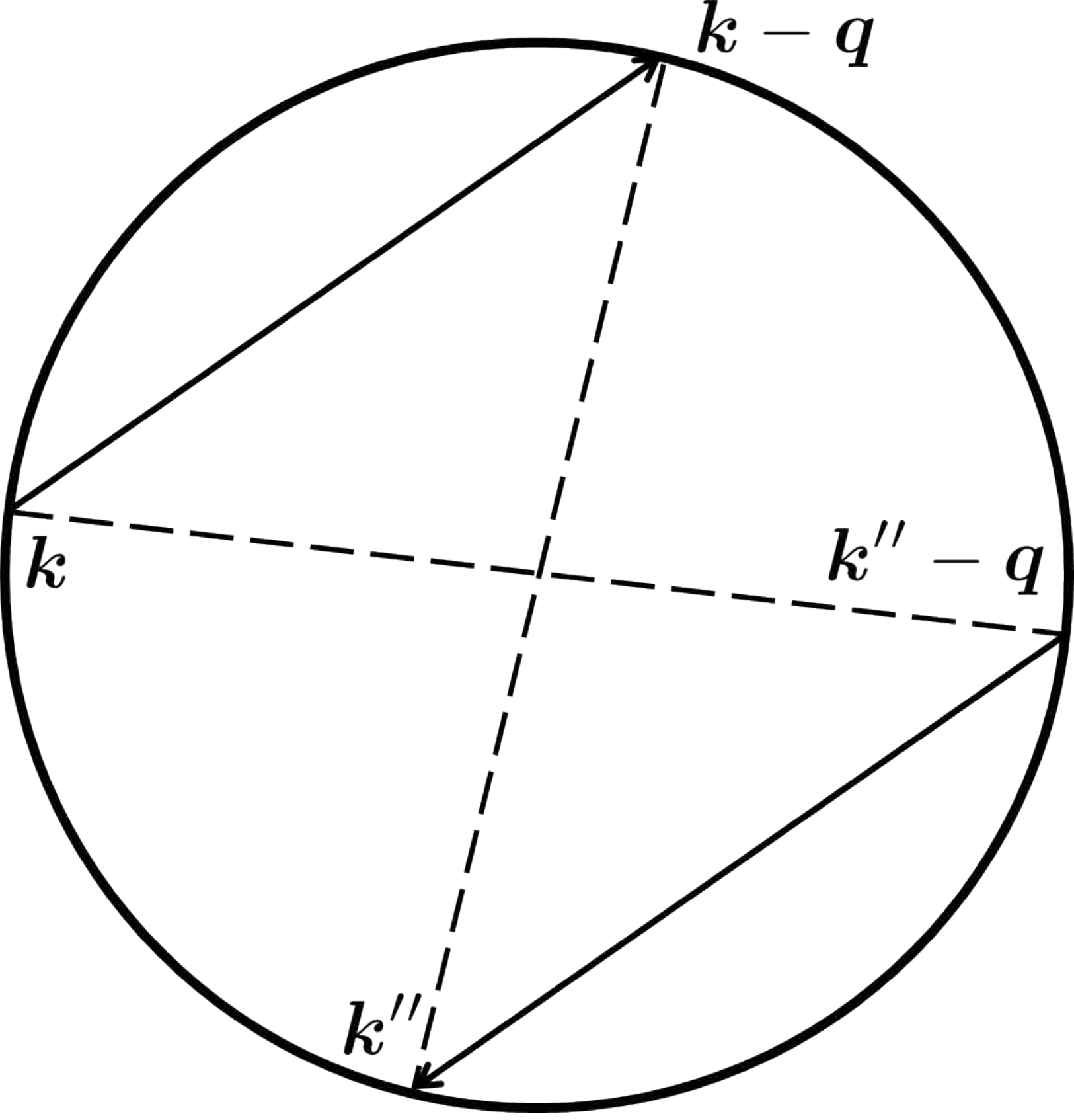}
\end{tabular}
\end{center}
\caption{
A pictorial representation of double particle-hole excitations that contribute, to lowest order in the strength of e-e interactions, to the quasiparticle decay rate calculated in Sect.~\ref{app:QP_lifetime}. Note that, since all the states involved in the scattering process live at the Fermi surface, the conservation of momentum constrains the initial states ${\bm k}$ and ${\bm k}'' - {\bm q}$ to be diametrically opposed. The same happens to the final states ${\bm k}-{\bm q}$ and ${\bm k}''$.
\label{fig:SM_two}}
\end{figure}

Although Eq.~(\ref{eq:SM_inverse_lifetime}) may look quite unfamiliar, shifting ${\bm k}' \to  {\bm k} - {\bm q}$ and ${\bm q}' \to -{\bm k}'' + {\bm k}$ we can recast is in the following Fermi-golden-rule form
\begin{eqnarray} \label{eq:SM_inverse_lifetime_2}
\frac{1}{\tau_{\rm qp}^{\rm ee}} 
&=&
\frac{4\pi}{3} (k_{\rm B} T)^2 \sum_{{\bm q}}
|W({\bm q},{\bar T}\varepsilon_{\rm F})|^2
\frac{\Im m \chi_{nn}({\bm q}, {\bar T}\varepsilon_{\rm F})}{{\bar T}\varepsilon_{\rm F}}
\nonumber\\
&\times&
\Im m \big[G^{({\rm R},\sigma)}_{+}({\bm k} - {\bm q},-{\bar T}\varepsilon_{\rm F})\big]
\frac{1+\cos(\varphi_{{\bm k}-{\bm q}}-\varphi_{{\bm k}})}{2}
~,
\nonumber\\
\end{eqnarray}
which describes, as shown in Fig.~\ref{fig:SM_two}, the decay (scattering) of a quasiparticle of momentum ${\bm k}$ to a state of momentum ${\bm k}-{\bm q}$  through the creation of an electron-hole pair of total momentum ${\bm q}$ obtained by transferring a particle of momentum ${\bm k}'' - {\bm q}$ to a state of momentum ${\bm k}'' $.  Such a process is encoded in the density-density response function $\Im m \chi_{nn}({\bm q}, {\bar T}\varepsilon_{\rm F})$ and is depicted in Fig.~\ref{fig:SM_two}. Notice that, since all the initial and final states are on the Fermi surface, the conservation of momentum implies that ${\bm k}$ and ${\bm k}'' - {\bm q}$ (and thus ${\bm k}-{\bm q}$ and ${\bm k}''$) are diametrically opposite.

%%%%%%%%%%%%%%%%%%%
%%%%%%%%%%%%%%%%%%%

To proceed further with the evaluation of Eq.~(\ref{eq:SM_inverse_lifetime_2}), we recall that, in the limit $\omega \to 0$,
\begin{eqnarray} \label{eq:SM_chi_nn_asymptotics}
\Re e \chi^{(0)}_{nn}(q,\omega) &=& -\nu(\varepsilon_{\rm F}) \Theta(v_{\rm F}^2 q^2-\omega^2) 
~,
\nonumber\\
\Im m \chi^{(0)}_{nn}(q,\omega) &=& -\omega \nu(\varepsilon_{\rm F}) \sqrt{1-\frac{q^2}{4 k_{\rm F}^2}} \frac{\Theta(v_{\rm F}^2 q^2-\omega^2)}{\sqrt{v_{\rm F}^2 q^2-\omega^2}}
\nonumber\\
&\times&
\Theta\big[v_{\rm F}^2 (2k_{\rm F}-q)^2 - \omega^2\big]
~,
\end{eqnarray}
where $\nu(\varepsilon_{\rm F}) = N_{\rm F} k_{\rm F}/(2\pi v_{\rm F})$ is the DOS at the Fermi energy.

The angular integration in Eq.~(\ref{eq:SM_inverse_lifetime_2}) can be easily performed with the help of the following formula~\cite{Polini_QP_lifetime}
\begin{eqnarray}
A(q,\omega) &=& -\frac{\pi}{v_{\rm F}} \int_0^{2\pi} d\varphi_{\bm q} \delta(|{\bm k}-{\bm q}| - k - \omega/v_{\rm F}) 
\nonumber\\
&\times&
\frac{1+\cos(\varphi_{{\bm k}-{\bm q}}+\varphi_{{\bm k}})}{2} \Bigg|_{k=k_{\rm F}}
\nonumber\\
&=&
-\frac{4\pi}{v_{\rm F}^2} \frac{k_{\rm F}-\omega/v_{\rm F}}{\sqrt{4 k_{\rm F}^2 q^2 - (q^2 - \omega^2/v_{\rm F}^2 + 2 k_{\rm F} \omega/v_{\rm F})^2}}
\nonumber\\
&\times&
\left( 1 - \frac{q^2-\omega^2/v_{\rm F}^2}{4 k_{\rm F}(k_{\rm F} - \omega/v_{\rm F})} \right)
\nonumber\\
&\times&
\Theta\left(1 - \left| \frac{q^2 - \omega^2/v_{\rm F}^2 + 2 k_{\rm F} \omega/v_{\rm F}}{2k_{\rm F}q} \right|\right)
~.
\end{eqnarray}
Eq.~(\ref{eq:SM_inverse_lifetime_2}) thus becomes
\begin{eqnarray} \label{eq:SM_inverse_lifetime_3}
\frac{1}{\tau_{\rm qp}^{\rm ee}} 
&=&
\frac{(k_{\rm B} T)^2}{3 \pi} \int_{k_{\rm F}{\bar T}}^{2 k_{\rm F}(1-{\bar T}/2)} dq~q
|W({\bm q},{\bar T}\varepsilon_{\rm F})|^2
\nonumber\\
&\times&
\frac{\Im m \chi_{nn}({\bm q}, {\bar T}\varepsilon_{\rm F})}{{\bar T}\varepsilon_{\rm F}}
A(q,-{\bar T}\varepsilon_{\rm F})
~,
\nonumber\\
\end{eqnarray}
where the boundaries of the integral are due to the Heaviside $\Theta$-functions on the right-hand side of Eq.~(\ref{eq:SM_chi_nn_asymptotics}).

\begin{figure}[t]
\begin{center}
\begin{tabular}{c}
\includegraphics[width=0.99\columnwidth]{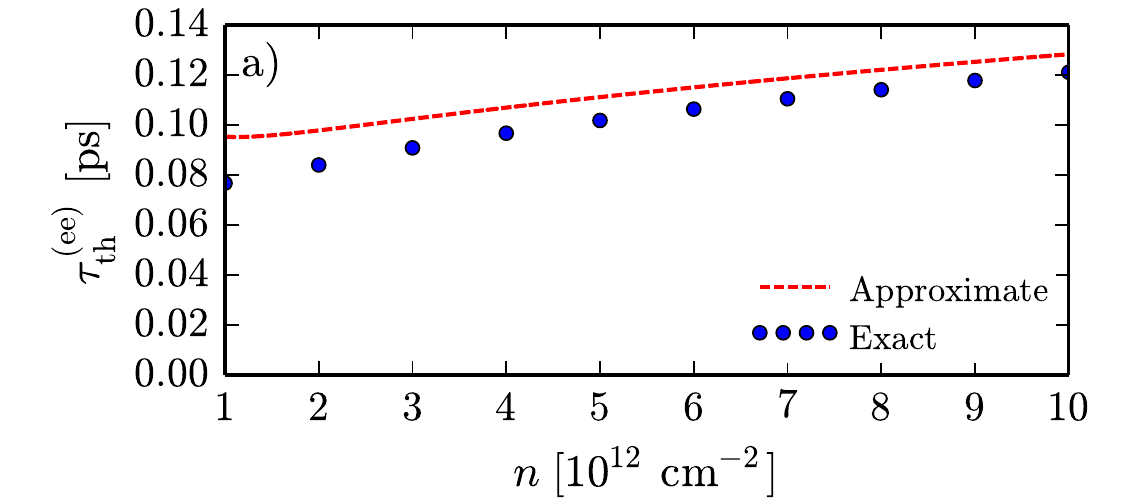}
\\
\includegraphics[width=0.99\columnwidth]{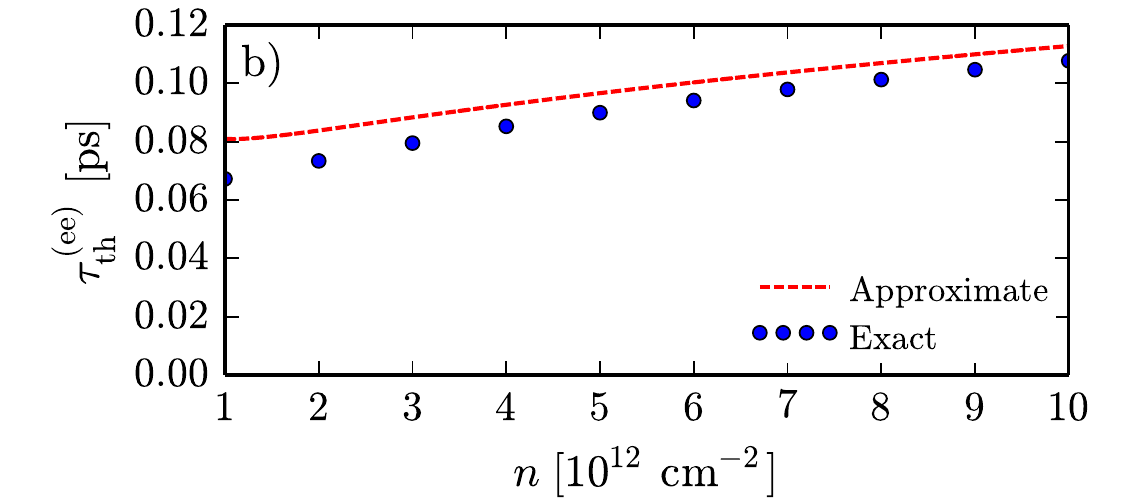}
\\
\includegraphics[width=0.99\columnwidth]{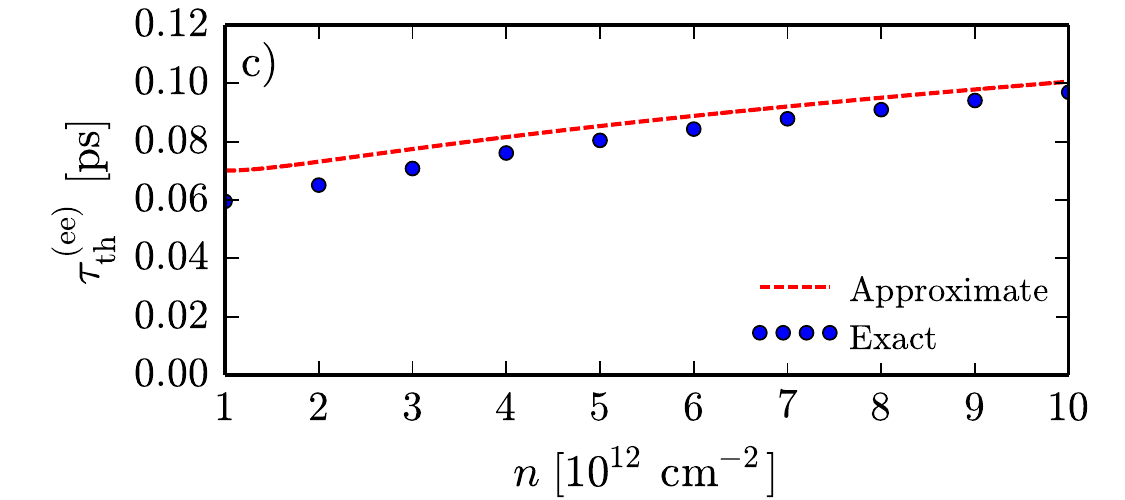}
\end{tabular}
\end{center}
\caption{
A comparison between the quasiparticle lifetime calculated from Eq.~(\ref{eq:SM_inverse_lifetime_3}) and the one computed in Ref.~\onlinecite{Polini_QP_lifetime}. In this figure the temperature is kept fixed at $T=300~{\rm K}$ and the quasiparticle lifetime is plot in units of ps as a function of the density (in units of $10^{12}~{\rm cm}^{-2}$). Panel a)-c) refer to the three values of the coupling constant $\alpha_{\rm ee}=0.5$, $\alpha_{\rm ee}=0.9$, and $\alpha_{\rm ee}=2.2$, respectively.
\label{fig:SM_three}}
\end{figure}

\begin{figure}[t]
\begin{center}
\begin{tabular}{c}
\includegraphics[width=0.99\columnwidth]{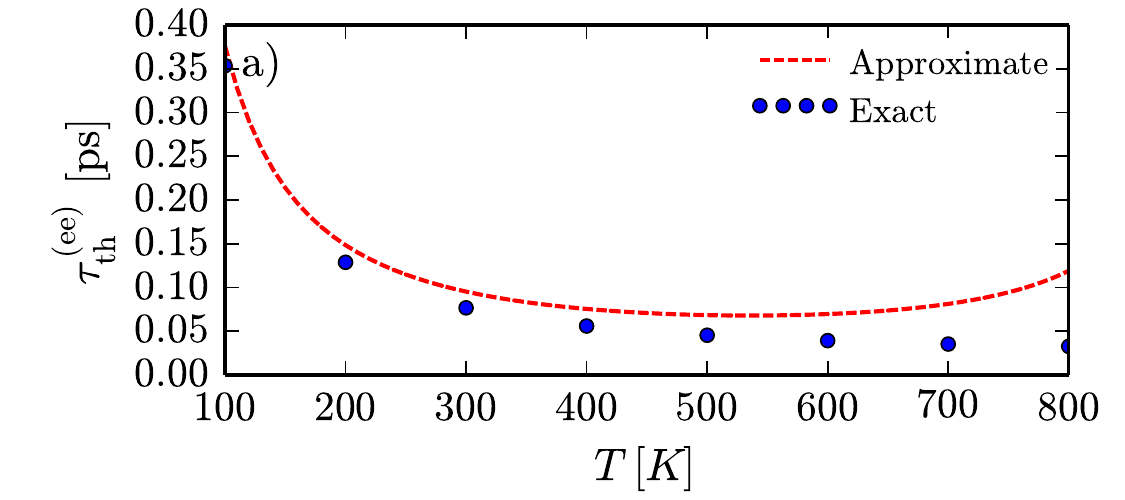}
\\
\includegraphics[width=0.99\columnwidth]{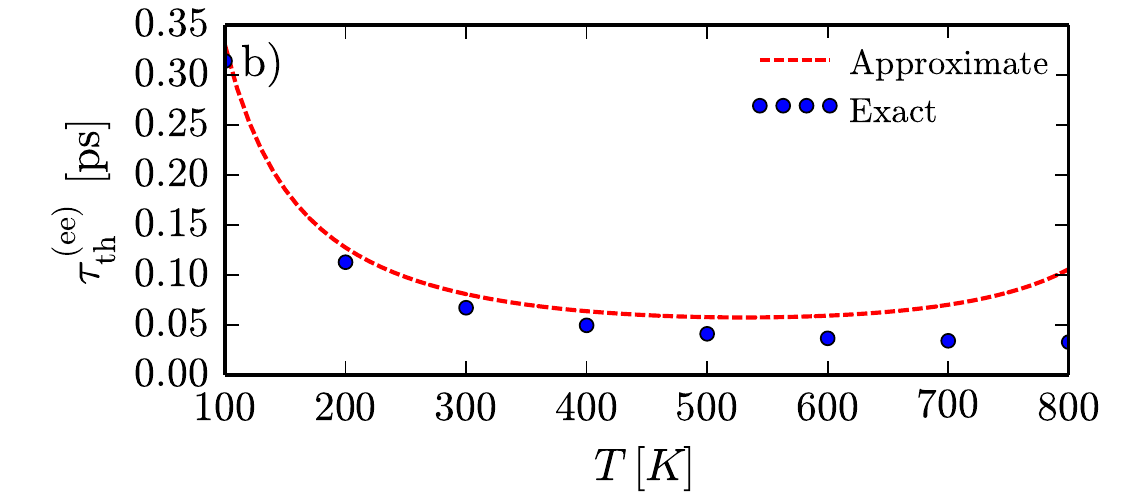}
\\
\includegraphics[width=0.99\columnwidth]{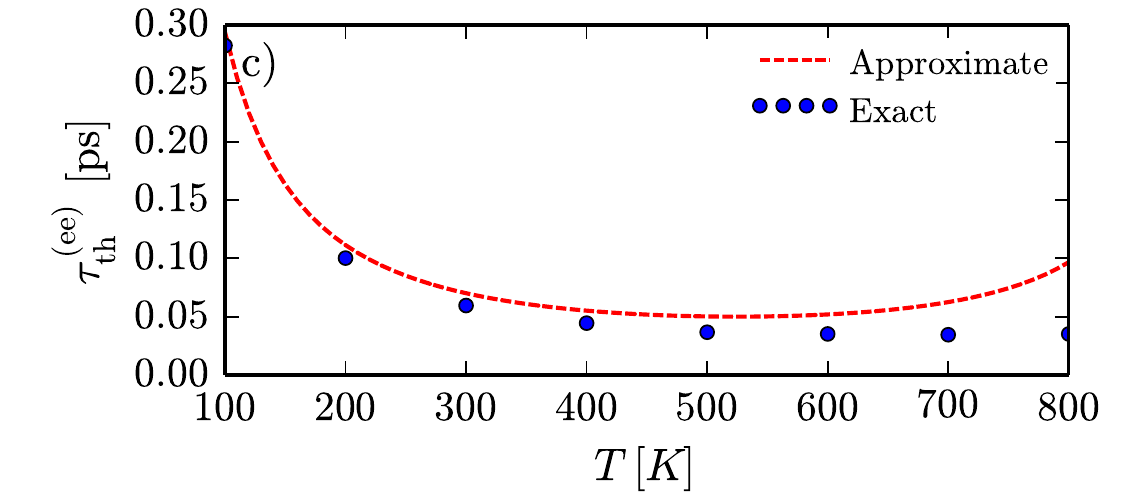}
\end{tabular}
\end{center}
\caption{
A comparison between the quasiparticle lifetime calculated from Eq.~(\ref{eq:SM_inverse_lifetime_3}) and the one computed in Ref.~\onlinecite{Polini_QP_lifetime}. In this figure the density is kept fixed at $n = 10^{12}~{\rm cm}^{-2}$ and the quasiparticle lifetime is plot in units of ps as a function of the temperature (measured in K). Panel a)-c) refer to the three values of the coupling constant $\alpha_{\rm ee}=0.5$, $\alpha_{\rm ee}=0.9$, and $\alpha_{\rm ee}=2.2$, respectively.
\label{fig:SM_four}}
\end{figure}

In Figs.~\ref{fig:SM_three}-\ref{fig:SM_four} we show a comparison between the quasiparticle lifetime calculated from Eq.~(\ref{eq:SM_inverse_lifetime_3}) and the ``exact'' one computed in Ref.~\onlinecite{Polini_QP_lifetime}. Note that the agreement is very good in the range of densities, temperatures and coupling constants explored in the main text of the Letter.  In comparison with Ref.~\onlinecite{Polini_QP_lifetime}, the key approximation of our calculation is encoded in Eq.~(\ref{eq:SM_epsilon_bar}), which allowed us to reduce the number of numerical integrations to be performed. 

%%%%%%%%%%%%%%%%%%%
%%%%%%%%%%%%%%%%%%%

%%%%%%%%%%%%%%%%%%%
%%%%%%%%%%%%%%%%%%%

\section{The derivation of the Bethe-Salpeter equation in the thermal channel}
\label{app:BS_thermal}
In this section we guide the reader through the long and complicated calculation of the vertex correction to the thermal conductivity. For this purpose, we start from the definition of the energy-current response function [Fig.~\ref{fig:SM_one}a)], which is given by
\begin{eqnarray} \label{eq:SM_chi_jj_def}
&& \!\!\!\!\!\!\!\!
\chi_{J_{\alpha}^{({\rm Q})} J_{\beta}^{({\rm Q})}} ({\bm q}, i\omega_m) = N_{\rm f} k_{\rm B} T \sum_{{\bm k}, \lambda, \lambda'} \sum_{\varepsilon_n} G_{\lambda}({\bm k}_-, i\varepsilon_n) 
\nonumber\\
&\times&
\Lambda^{(0)}_{\lambda\lambda',\alpha}({\bm k}_-,i\varepsilon_n,{\bm k}_+, i\varepsilon_n + i\omega_m) G_{\lambda'}({\bm k}_+, i\varepsilon_n + i\omega_m)
\nonumber\\
&\times&
\Lambda_{\lambda'\lambda,\beta} ({\bm k}_+, i\varepsilon_n + i\omega_m,{\bm k}_-,i\varepsilon_n)
~.
\end{eqnarray}
Here $\alpha,\beta=x,y$ are Cartesian indices, $G_{\lambda}({\bm k}_, i\varepsilon_n)$ is the Green's function dressed by self-energy insertions [see Fig.~\ref{fig:SM_one}b) and Sect.~\ref{app:QP_lifetime}], and
\begin{equation} \label{SM_BareVertex}
\Lambda^{(0)}_{\lambda\lambda',\beta}({\bm k}_-,i\varepsilon_n,{\bm k}_+,i\varepsilon_n+i\omega_m) =
(\varepsilon_n + \omega_m/2) J^{\lambda\lambda'}_{{\bm k}_-,{\bm k}_+,\beta}
~, 
\end{equation}
is the bare energy-current vertex. The number-current vertex is defined in Eq.~(\ref{eq:SM_J_element}). Finally, the dressed vertex satisfies the Bethe-Salpeter equation [Fig.~\ref{fig:SM_one}c)]
\begin{eqnarray} \label{eq:SM_Lambda_def}
&& \!\!\!\!\!\!\!\!
\Lambda_{\lambda'\lambda,\beta} ({\bm k}_+, i\varepsilon_n + i\omega_m,{\bm k}_-,i\varepsilon_n) = 
\nonumber\\
&\times&
\Lambda^{(0)}_{\lambda'\lambda,\beta} ({\bm k}_+, i\varepsilon_n + i\omega_m,{\bm k}_-,i\varepsilon_n)
\nonumber\\
&+&
\sum_{i=1,\ldots,3} \Lambda^{(i)}_{\lambda'\lambda,\beta} ({\bm k}_+, i\varepsilon_n + i\omega_m,{\bm k}_-,i\varepsilon_n)
~.
\end{eqnarray}
The functions $\Lambda^{(i)}_{\lambda'\lambda,\beta} ({\bm k}_+, i\varepsilon_n + i\omega_m,{\bm k}_-,i\varepsilon_n)$ (with $i=1,\ldots,3$) refer to the last three diagrams on the right-hand side of Fig.~\ref{fig:SM_one}c), and are themselves expressed in terms of the full vertex $\Lambda$, yielding a self-consistent equation for the latter.  Their explicit expressions are
\begin{eqnarray} \label{eq:SM_Lambda_1_2}
&& \!\!\!\!\!\!\!\!
\Lambda^{(1,2)}_{\lambda'\lambda,\beta} ({\bm k}_+, i\varepsilon_n + i\omega_m,{\bm k}_-,i\varepsilon_n) =
-k_{\rm B} T \sum_{{\bm k}',\varepsilon_{n'}} \sum_{\mu,\mu'} 
\nonumber\\
&\times& 
W^{(1,2)}_{\lambda\lambda'\mu\mu'}({\bm k}',{\bm k},i\varepsilon_{n'}-i\varepsilon_n)
G_{\mu'}({\bm k}'_+,i\varepsilon_{n'}+i\omega_m)
\nonumber\\
&\times&
G_{\mu}({\bm k}'_-,i\varepsilon_{n'})
\Lambda_{\mu'\mu,\beta} ({\bm k}'_+, i\varepsilon_{n'} + i\omega_m,{\bm k}'_-,i\varepsilon_{n'})
~,
\nonumber\\
\end{eqnarray}
and
\begin{eqnarray} \label{eq:SM_Lambda_3}
&& \!\!\!\!\!\!\!\!
\Lambda^{(3)}_{\lambda'\lambda,\beta} ({\bm k}_+, i\varepsilon_n + i\omega_m,{\bm k}_-,i\varepsilon_n) =
-k_{\rm B} T \sum_{{\bm k}',\varepsilon_{n'}} \sum_{\mu,\mu'}
\nonumber\\
&\times&
W^{(3)}_{\lambda\lambda'\mu\mu'}({\bm k}', {\bm k}, i\varepsilon_{n'}+i\varepsilon_{n} +i\omega_m)
G_{\mu'}({\bm k}'_+,i\varepsilon_{n'}+i\omega_m) 
\nonumber\\
&\times&
G_{\mu}({\bm k}'_-,i\varepsilon_{n'})
\Lambda_{\mu'\mu,\beta} ({\bm k}'_+, i\varepsilon_{n'} + i\omega_m,{\bm k}'_-,i\varepsilon_{n'})
~.
\nonumber\\
\end{eqnarray}
Here we have defined
\begin{eqnarray} \label{eq:SM_W_1}
W^{(1)}_{\lambda\lambda'\mu\mu'} ({\bm k}',{\bm k},i\omega_m) &=& W({\bm k}-{\bm k}',i\omega_m) 
{\cal D}_{\lambda'\mu'}({\bm k}_+,{\bm k}'_+) 
\nonumber\\
&\times&
{\cal D}_{\mu\lambda}({\bm k}'_-,{\bm k}_-)
~,
\end{eqnarray}
and
\begin{eqnarray} \label{eq:SM_W_2}
&& \!\!\!\!\!\!\!\!
W^{(2)}_{\lambda\lambda'\mu\mu'} ({\bm k}',{\bm k},i\varepsilon_{n'}-i\varepsilon_n) =  \frac{N_{\rm f}}{\beta} \sum_{\substack{ {\bm q}',\omega_{m'}\\ \lambda'',\mu'' }} W({\bm q}',i\omega_{m'}) 
\nonumber\\
&\times&
W({\bm q}'-{\bm q},i\omega_{m'}-i\omega_m) 
{\cal D}_{\lambda'\lambda''}({\bm k}_+,{\bm k}_+-{\bm q}')
\nonumber\\
&\times&
{\cal D}_{\lambda''\lambda}({\bm k}_+-{\bm q}',{\bm k}_-)
{\cal D}_{\mu\mu''}({\bm k}'_-,{\bm k}'_+-{\bm q}')
\nonumber\\
&\times&
{\cal D}_{\mu''\mu'}({\bm k}'_+-{\bm q}',{\bm k}'_+)
G_{\lambda''}({\bm k}_+-{\bm q}',i\varepsilon_n+i\omega_m -i\omega_{m'})
\nonumber\\
&\times&
G_{\mu''}({\bm k}'_+-{\bm q}',i\varepsilon_{n'}+i\omega_m -i\omega_{m'})
~,
\end{eqnarray}
and finally
\begin{eqnarray} \label{eq:SM_W_3}
&& \!\!\!\!\!\!\!\!
W^{(3)}_{\lambda\lambda'\mu\mu'} ({\bm k}',{\bm k},i\varepsilon_{n'}+i\varepsilon_n+i\omega_m) =  \frac{N_{\rm f}}{\beta} \sum_{\substack{ {\bm q}',\omega_{m'}\\ \lambda'',\mu''} }
W({\bm q}',i\omega_{m'})
\nonumber\\
&\times&
W({\bm q}'-{\bm q},i\omega_{m'}-i\omega_m) 
{\cal D}_{\lambda\lambda''}({\bm k}_-,{\bm k}_-+{\bm q}')
\nonumber\\
&\times&
{\cal D}_{\lambda''\lambda'}({\bm k}_-+{\bm q}',{\bm k}_+)
{\cal D}_{\mu\mu''}({\bm k}'_-,{\bm k}'_+-{\bm q}')
\nonumber\\
&\times&
{\cal D}_{\mu''\mu'}({\bm k}'_+-{\bm q}',{\bm k}'_+)
G_{\lambda''}({\bm k}_-+{\bm q}',i\varepsilon_n +i\omega_{m'})
\nonumber\\
&\times&
G_{\mu''}({\bm k}'_+-{\bm q}',i\varepsilon_{n'}+i\omega_m -i\omega_{m'})
~.
\end{eqnarray}

\subsection{The analytical continuation of Eq.~(\ref{eq:SM_chi_jj_def})}
The first step of this  derivation consists in analytically continuing Eq.~(\ref{eq:SM_chi_jj_def}) to real frequencies by means of the standard replacement $i\omega_m \to \omega+i\eta$. We define the function $f(i\varepsilon,i\varepsilon+i\omega)$ such that
\begin{eqnarray} \label{eq:SM_chi_jj_f}
\chi_{j^{({\rm Q})}_{\alpha} j^{({\rm Q})}_{\beta}} ({\bm q}, i\omega_m) &\equiv& k_{\rm B} T \sum_{\varepsilon_n} f(i\varepsilon_n, i\varepsilon_n+i\omega_m)
\nonumber\\
&=& -\oint_{\cal C} \frac{dz}{2\pi i} n_{\rm F}(z) f(z,z+i\omega_m)
~,
\nonumber\\
\end{eqnarray}
where we suppress for brevity the dependence on momenta. The contour ${\cal C}$ in Eq.~(\ref{eq:SM_chi_jj_f}) is chosen in such a way as to encircle the poles of $n_{\rm F}(z)$ and to exclude the branch cuts of $f(z,z+i\omega_m)$, which occur for $\Im m (z) = 0$ and $\Im m (z+i\omega_m) = 0$. Taking the limit $i\omega_m \to \omega + i\eta$. We get
\begin{eqnarray} \label{eq:SM_chi_jj_final}
&& \!\!\!\!\!\!\!\!
\chi_{j^{({\rm Q})}_{\alpha} j^{({\rm Q})}_{\beta}} ({\bm q}, \omega) = -\int \frac{d\varepsilon}{2\pi i} \Big\{\big[n_{\rm F} (\varepsilon+\omega) - n_{\rm F} (\varepsilon)\big] 
\nonumber\\
&\times&
\big[f(\varepsilon_-, \varepsilon_++\omega) - f(\varepsilon_-, \varepsilon_-+\omega)\big]
\nonumber\\
&+&
n_{\rm F}(\varepsilon) \big[f(\varepsilon_+, \varepsilon_++\omega) - f(\varepsilon_-, \varepsilon_-+\omega)\big] \Big\}
~.
\end{eqnarray}

Note that the square brackets in the last line of Eq.~(\ref{eq:SM_chi_jj_final}) contain a purely imaginary quantity, which (being divided by the imaginary unit) gives a purely real contribution to $\chi_{j^{({\rm Q})}_{\alpha} j^{({\rm Q})}_{\beta}} ({\bm q}, \omega)$. Note also that $f(\varepsilon_-, \varepsilon_++\omega)$ contains the product of a retarded and an advanced Green's function, whereas in $f(\varepsilon_-, \varepsilon_-+\omega)$ and $f(\varepsilon_+, \varepsilon_++\omega)$ both Green's functions are either advanced or retarded. The last two functions [$f(\varepsilon_-, \varepsilon_-+\omega)$ and $f(\varepsilon_+, \varepsilon_++\omega)$] have all the poles on the same half of the complex plane. 
Thus, $f(\varepsilon_-, \varepsilon_++\omega)$ gives the dominant contribution in the limit $\varepsilon_{\rm F} \tau_{\rm qp}^{\rm ee} \gg 1$. In what follows we retain only this term. Eq.~(\ref{eq:SM_chi_jj_final}) thus becomes
\begin{eqnarray} \label{eq:SM_chi_jj_final_omega_finite}
&& \!\!\!\!\!\!\!\!
\chi_{j^{({\rm Q})}_{\alpha} j^{({\rm Q})}_{\beta}} ({\bm q}, \omega) = - N_{\rm f} \sum_{{\bm k}, \lambda, \lambda'} \int \frac{d\varepsilon}{2\pi i} \big[n_{\rm F} (\varepsilon+\omega) - n_{\rm F} (\varepsilon)\big]
\nonumber\\
&\times&
G^{({\rm A})}_{\lambda}({\bm k}_-, \varepsilon) \Lambda^{(0)}_{\lambda\lambda',\alpha}({\bm k}_-,\varepsilon_-,{\bm k}_+, \varepsilon_+ + \omega)
G^{({\rm R})}_{\lambda'}({\bm k}_+, \varepsilon + \omega)
\nonumber\\
&\times&
\Lambda_{\lambda'\lambda,\beta} ({\bm k}_+, \varepsilon_+ + \omega,{\bm k}_-,\varepsilon_-)
~.
\end{eqnarray}

\subsection{The analytical continuation of the Bethe-Salpeter equation}
The second step consists in analytically continuing the three contributions $\Lambda^{(i,\sigma\sigma')}_{\lambda'\lambda,\beta} ({\bm k}_+, i\varepsilon_n + i\omega_m,{\bm k}_-,i\varepsilon_n)$ ($i=1,\ldots,3$), defined in Eqs.~(\ref{eq:SM_Lambda_1_2})-(\ref{eq:SM_W_3}) to real frequencies.  In the Fermi-liquid regime we consider only the dominant contribution to the dressed vertex, to be used in combination with the product of the retarded and advanced Green's functions that appears in Eq.~(\ref{eq:SM_chi_jj_final_omega_finite}). From this we see that the analytic continuation of $\Lambda^{(i,\sigma\sigma')}_{\lambda'\lambda,\beta} ({\bm k}_+, i\varepsilon_n + i\omega_m,{\bm k}_-,i\varepsilon_n)$ is done with the prescriptions $i\omega_m\to \omega_+$, $i\varepsilon_n\to \varepsilon_-$, $i\varepsilon_n+i\omega_m \to \varepsilon_++\omega$.

\subsubsection{Analytical continuation of Eq.~(\ref{eq:SM_Lambda_1_2})}
We define the function $g(i\varepsilon_{n'},i\varepsilon_{n'} + i\omega_m,i\varepsilon_{n'}-i\varepsilon_n)$ such that
\begin{eqnarray} \label{eq:SM_Lambda_1_2_f}
&& \!\!\!\!\!\!\!\!
\Lambda^{(1,2)} ({\bm k}_+, i\varepsilon_n + i\omega_m,{\bm k}_-,i\varepsilon_n) 
\nonumber\\
&\equiv&
-k_{\rm B} T \sum_{\varepsilon_{n'}} g(i\varepsilon_{n'},i\varepsilon_{n'} + i\omega_m,i\varepsilon_{n'}-i\varepsilon_n)
\nonumber\\
&=&
\oint \frac{dz}{2\pi i} n_{\rm F}(z) g(z,z + i\omega_m,z-i\varepsilon_n)
~.
\end{eqnarray}
Here and in what follows we suppress for brevity all the band and spatial indices of the dressed vertex. As usual, we transform the sum over the poles of $n_{\rm F}(z)$ in an integration over the branch cuts of $g(z,z + i\omega_m,z-i\varepsilon_n)$. We then perform the analytic continuations with the prescription $i\omega_m\to \omega_+$, $i\varepsilon_n\to \varepsilon_-$, $i\varepsilon_n+i\omega_m \to \varepsilon_++\omega$. After some lengthy but straightforward algebra we get
\begin{eqnarray} \label{eq:SM_Lambda_1_2_f_BC}
&& \!\!\!\!\!\!\!\!
\Lambda^{(1,2)} ({\bm k}_+, \varepsilon_+ + \omega,{\bm k}_-,\varepsilon_-) =
\int \frac{d\varepsilon'}{2\pi i} \Big\{ 
n_{\rm F}(\varepsilon') 
\nonumber\\
&\times&
\big[ g(\varepsilon'_+,\varepsilon'_+ + \omega,\varepsilon'_+-\varepsilon) - g(\varepsilon'_-,\varepsilon'_+ + \omega,\varepsilon'_+-\varepsilon)\big]
\nonumber\\
&+&
n_{\rm F}(\varepsilon') \big[ g(\varepsilon'_--\omega,\varepsilon'_+ ,\varepsilon'_--\varepsilon-\omega) 
\nonumber\\
&-& g(\varepsilon'_--\omega,\varepsilon'_-,\varepsilon'_--\varepsilon-\omega)\big]
\nonumber\\
&-&
n_{\rm B}(\varepsilon')
\big[ g(\varepsilon'_-+\varepsilon,\varepsilon'_+ +\varepsilon + \omega,\varepsilon'_+) 
\nonumber\\
&-&
g(\varepsilon'_-+\varepsilon,\varepsilon'_+ + \varepsilon+\omega,\varepsilon'_-)\big]
\Big\}
~.
\end{eqnarray}
We now shift $\varepsilon'\to \varepsilon'+\omega$ in the third and fourth lines of Eq.~(\ref{eq:SM_Lambda_1_2_f_BC}). We also note that we can safely take the limit $\omega \to 0$ in $n_{\rm F}(\varepsilon'+\omega)$. Notice also that $g(\varepsilon'_+,\varepsilon'_+ + \omega,\varepsilon'_+-\varepsilon)$ and $g(\varepsilon'_--\omega,\varepsilon'_-,\varepsilon'_--\varepsilon-\omega)$ have the poles on the same half of the complex plane, and therefore can be neglected in the limit $\varepsilon_{\rm F} \tau_{\rm qp}^{\rm ee} \gg 1$. Shifting $\varepsilon'\to \varepsilon'+\varepsilon$ in the last two lines of Eq.~(\ref{eq:SM_Lambda_1_2_f_BC}) we readily obtain
\begin{eqnarray} \label{eq:SM_Lambda_1_2_f_final}
&& \!\!\!\!\!\!\!\!
\Lambda^{(1,2)} ({\bm k}_+, \varepsilon_++\omega,{\bm k}_-,\varepsilon_-) =
\int \frac{d\varepsilon'}{2\pi i}
\big[ n_{\rm B}(\varepsilon'-\varepsilon) + n_{\rm F}(\varepsilon')\big]
\nonumber\\
&\times&
\big[ g(\varepsilon'_-,\varepsilon'_++\omega ,\varepsilon'_--\varepsilon) - g(\varepsilon'_-,\varepsilon'_++\omega,\varepsilon'_+-\varepsilon) \big]
\nonumber\\
&=& 
\sum_{{\bm k}'} \sum_{\mu,\mu'} \int \frac{d\varepsilon'}{2\pi i} \big[n_{\rm F}(\varepsilon') + n_{\rm B}(\varepsilon'-\varepsilon)\big]
\nonumber\\
&\times&
\Big[ W^{(1,2)}_{\lambda\lambda'\mu\mu'}({\bm k}',{\bm k},\varepsilon'_--\varepsilon) - W^{(1,2,\sigma\sigma'')}_{\lambda\lambda'\mu\mu'}({\bm k}',{\bm k},\varepsilon'_+-\varepsilon) \Big]
\nonumber\\
&\times&
G^{({\rm R})}_{\mu'}({\bm k}'_+,\varepsilon'+\omega) G^{({\rm A})}_{\mu}({\bm k}'_-,\varepsilon')
\nonumber\\
&\times&
\Lambda_{\mu'\mu,\beta} ({\bm k}'_+, \varepsilon'_++\omega, {\bm k}'_-,\varepsilon'_-)
~.
\end{eqnarray}
It remains to determine $W^{(1,2)}_{\lambda\lambda'\mu\mu'}({\bm k}',{\bm k},\varepsilon'_\pm-\varepsilon)$. Eq.~(\ref{eq:SM_Lambda_1_2_f_final}) implies that we have to analytically continue the functions $W^{(1,2)}$ with the prescription $i\varepsilon_{n'} \to \varepsilon'_-$ and $i\varepsilon_{n'} + i\omega_m \to \varepsilon'_+ + \omega$.

\subsubsection{The analytical continuation of Eq.~(\ref{eq:SM_W_1})}
We now perform the analytical continuation of Eq.~(\ref{eq:SM_W_1}) with the prescription $i\omega_m\to \omega_+$, $i\varepsilon_n\to \varepsilon_-$, $i\varepsilon_n+i\omega_m \to \varepsilon_++\omega$, $i\varepsilon_{n'} \to \varepsilon'_-$, and $i\varepsilon_{n'} + i\omega_m \to \varepsilon'_+ + \omega$. As shown in Eq.~(\ref{eq:SM_Lambda_1_2_f_final}), we need to calculate the difference $W^{(1)}_{\lambda\lambda'\mu\mu'}({\bm k},{\bm k'},\varepsilon_-'-\varepsilon) - W^{(1)}_{\lambda\lambda'\mu\mu'}({\bm k},{\bm k'},\varepsilon_+'-\varepsilon)$, which reads
\begin{eqnarray} \label{eq:SM_W_1_A-R}
&& \!\!\!\!\!\!\!\!
W^{(1)}_{\lambda\lambda'\mu\mu'}({\bm k},{\bm k'},\varepsilon_-'-\varepsilon) - W^{(1)}_{\lambda\lambda'\mu\mu'}({\bm k},{\bm k'},\varepsilon_+'-\varepsilon) = 
\nonumber\\
&& \!\!\!\!\!\!\!\!
\big[W({\bm k}-{\bm k}',\varepsilon_-'-\varepsilon) - W({\bm k}-{\bm k}',\varepsilon_+'-\varepsilon)\big]
\nonumber\\
&\times&
{\cal D}_{\lambda'\mu'}({\bm k}_+,{\bm k}'_+) {\cal D}_{\mu\lambda}({\bm k}'_-,{\bm k}_-)
\nonumber\\
&=&
- 2 i \Im m \big[W({\bm k}-{\bm k}',\varepsilon_+'-\varepsilon)\big]
\nonumber\\
&\times&
{\cal D}_{\lambda'\mu'}({\bm k}_+,{\bm k}'_+) {\cal D}_{\mu\lambda}({\bm k}'_-,{\bm k}_-)
\nonumber\\
&=& \!\!\!
-2 i |W({\bm k}-{\bm k}',\varepsilon'-\varepsilon)|^2 \Im m \big[\chi_{nn}^{(0)}({\bm k}-{\bm k}',\varepsilon_+'-\varepsilon)\big]
\nonumber\\
&\times&
{\cal D}_{\lambda'\mu'}({\bm k}_+,{\bm k}'_+) {\cal D}_{\mu\lambda}({\bm k}'_-,{\bm k}_-)
~.
\end{eqnarray}
The density-density response function was derived in Sect.~\ref{app:QP_lifetime}. Putting Eq.~(\ref{eq:SM_app_Im_chi_nn}) into Eq.~(\ref{eq:SM_W_1_A-R}) we finally find
\begin{eqnarray} \label{eq:SM_W_1_A-R_2}
&& \!\!\!\!\!\!\!\!
W^{(1)}_{\lambda\lambda'\mu\mu'}({\bm k},{\bm k'},\varepsilon_-'-\varepsilon) - W^{(1)}_{\lambda\lambda'\mu\mu'}({\bm k},{\bm k'},\varepsilon_+'-\varepsilon)  =
4 N_{\rm f} 
\nonumber\\
&\times&
|W({\bm k}-{\bm k}',\varepsilon'-\varepsilon)|^2
\int \frac{d\omega'}{2\pi i} \big[n_{\rm F} (\omega'+\varepsilon') - n_{\rm F} (\omega'+\varepsilon)\big] 
\nonumber\\
&\times&
\sum_{{\bm q}',\lambda'',\mu''}
\Im m \Big[G^{({\rm R})}_{\lambda''}({\bm q}'-{\bm k},\omega'+\varepsilon)\Big]
\nonumber\\
&\times&
\Im m\Big[G^{({\rm R})}_{\mu''}({\bm q'}-{\bm k}', \omega'+\varepsilon')\Big]
{\cal D}_{\lambda'\mu'}({\bm k}_+,{\bm k}'_+) 
{\cal D}_{\mu\lambda}({\bm k}'_-,{\bm k}_-)
\nonumber\\
&\times&
{\cal D}_{\lambda''\mu''}({\bm q}'-{\bm k},{\bm q}'-{\bm k}') {\cal D}_{\mu''\lambda''}({\bm q}'-{\bm k}',{\bm q}'-{\bm k}) 
~.
\nonumber\\
\end{eqnarray}

\subsubsection{The analytical continuation of Eq.~(\ref{eq:SM_W_2})}
We now turn to the analytical continuation of Eq.~(\ref{eq:SM_W_2}) with the prescription $i\omega_m\to \omega_+$, $i\varepsilon_n\to \varepsilon_-$, $i\varepsilon_n+i\omega_m \to \varepsilon_++\omega$, $i\varepsilon_{n'} \to \varepsilon'_-$, and $i\varepsilon_{n'} + i\omega_m \to \varepsilon'_+ + \omega$. This time we define 
\begin{eqnarray} \label{eq:SM_W_1_2_f}
&& \!\!\!\!\!\!\!\!
W^{(2)}_{\lambda\lambda'\mu\mu'} ({\bm k}',{\bm k},i\varepsilon_{n'}-i\varepsilon_n) \equiv 
\oint \frac{dz}{2\pi i} n_{\rm B}(z) 
\nonumber\\
&\times&
w_2 (z, z-i\omega_m,i\varepsilon_{n}+i\omega_m -z, i\varepsilon_{n'}+i\omega_m -z)
~.
\nonumber\\
\end{eqnarray}
Integrating over the branch cuts of $w_2 (z, z-i\omega_m,i\varepsilon_{n}+i\omega_m -z, i\varepsilon_{n'}+i\omega_m -z)$ and performing the analytical continuations as stated before Eq.~(\ref{eq:SM_W_1_2_f}) we get
\begin{eqnarray} \label{eq:SM_W_1_2_f_BC}
&& \!\!\!\!\!\!\!\!
W^{(2)}_{\lambda\lambda'\mu\mu'} ({\bm k}',{\bm k},\varepsilon'_\pm-\varepsilon) =
\int \frac{d\omega'}{2\pi i} \Big\{
n_{\rm B}(\omega') 
\nonumber\\
&\times&
\big[w_2 (\omega'_+, \omega'_--\omega,\varepsilon + \omega -\omega'_-, \varepsilon'+\omega -\omega'_-) 
\nonumber\\
&-&
w_2 (\omega'_-, \omega'_--\omega,\varepsilon + \omega -\omega'_-, \varepsilon'+\omega -\omega'_-)\big]
\nonumber\\
&+&
n_{\rm B}(\omega') \big[w_2 (\omega'_++\omega, \omega'_+,\varepsilon -\omega'_+, \varepsilon' -\omega'_+) 
\nonumber\\
&-&
w_2 (\omega'_++\omega, \omega'_-,\varepsilon -\omega'_+, \varepsilon' -\omega'_+) \big]
\nonumber\\
&-&
n_{\rm F}(\omega') \big[w_2 (\omega'_++\varepsilon+\omega, \omega'_-+\varepsilon, -\omega'_+, \varepsilon'-\varepsilon -\omega'_\mp) 
\nonumber\\
&-&
w_2 (\omega'_++\varepsilon+\omega, \omega'_-+\varepsilon, -\omega'_-, \varepsilon'-\varepsilon -\omega'_\mp) \big]
\nonumber\\
&-&
n_{\rm F}(\omega') \big[w_2 (\omega'_++\varepsilon'+\omega, \omega'_-+\varepsilon', \varepsilon-\varepsilon' -\omega'_\pm, -\omega'_+) 
\nonumber\\
&-&
w_2 (\omega'_++\varepsilon'+\omega, \omega'_-+\varepsilon', \varepsilon-\varepsilon' -\omega'_\pm, -\omega'_-) \big]
\Big\}
~.
\nonumber\\
\end{eqnarray}
Note that the terms on the r.h.s. of Eq.~(\ref{eq:SM_W_1_2_f_BC}) proportional to $n_{\rm B}(\omega')$ are identical in both $W^{(2)}_{\lambda\lambda'\mu\mu'} ({\bm k}',{\bm k},\varepsilon'_\pm-\varepsilon)$ and thus vanish when the difference is taken. We thus neglect them in what follows. Eq.~(\ref{eq:SM_W_1_2_f_BC}) thus reduces to
\begin{eqnarray} \label{eq:SM_W_1_2_f_final}
&& \!\!\!\!\!\!\!\!
W^{(2)}_{\lambda\lambda'\mu\mu'} ({\bm k}',{\bm k},\varepsilon'_\pm-\varepsilon) =
-\int \frac{d\omega'}{2\pi i} \Big\{
n_{\rm F}(\omega'-\varepsilon-\omega)
\nonumber\\
&\times&
\big[w_2 (\omega'_+, \omega'_--\omega, \varepsilon+\omega-\omega'_+, \varepsilon' +\omega-\omega'_\mp) 
\nonumber\\
&-&
w_2 (\omega'_+, \omega'_--\omega, \varepsilon+\omega-\omega'_-, \varepsilon'+\omega-\omega'_\mp) \big]
\nonumber\\
&+&
n_{\rm F}(\omega'-\varepsilon'-\omega)
\nonumber\\
&\times&
\big[w_2 (\omega'_++, \omega'_--\omega, \varepsilon +\omega-\omega'_\pm, \varepsilon'+\omega-\omega'_+) 
\nonumber\\
&-&
w_2 (\omega'_+, \omega'_--\omega, \varepsilon +\omega-\omega'_\pm, \varepsilon'+\omega-\omega'_-) \big]
\Big\}
~.
\nonumber\\
\end{eqnarray}
Finally,
\begin{eqnarray} \label{eq:SM_W_1_2_A-R}
&& \!\!\!\!\!\!\!\!
W^{(2)}_{\lambda\lambda'\mu\mu'} (\varepsilon'_--\varepsilon) - W^{(2)}_{\lambda\lambda'\mu\mu'} (\varepsilon'_+-\varepsilon) = 
\int \frac{d\omega'}{2\pi i} 
\nonumber\\
&\times&
\big[n_{\rm F}(\omega'-\varepsilon-\omega) - n_{\rm F}(\omega'-\varepsilon'-\omega)\big]
\nonumber\\
&\times&
\big[
w_2 (\omega'_+, \omega'_--\omega, \varepsilon+\omega-\omega'_+, \varepsilon'+\omega -\omega'_-) 
\nonumber\\
&-&
w_2 (\omega'_+, \omega'_--\omega, \varepsilon+\omega-\omega'_-, \varepsilon'+\omega-\omega'_-) 
\nonumber\\
&-&
w_2 (\omega'_+, \omega'_--\omega, \varepsilon+\omega-\omega'_+, \varepsilon'+\omega -\omega'_+) 
\nonumber\\
&+&
w_2 (\omega'_+, \omega'_--\omega, \varepsilon+\omega-\omega'_-, \varepsilon'+\omega-\omega'_+)\big]
\nonumber\\
&=&
4 N_{\rm f} \sum_{{\bm q}',\lambda'',\mu''} \int \frac{d\omega'}{2\pi i} 
W({\bm q}',\omega'_+) W({\bm q}',\omega'_--\omega)
\nonumber\\
&\times&
\big[n_{\rm F}(\omega'-\varepsilon-\omega) - n_{\rm F}(\omega'-\varepsilon'-\omega)\big]
\nonumber\\
&\times&
\Im m\big[ G^{({\rm R})}_{\lambda''}({\bm k}_+-{\bm q}',\varepsilon+\omega-\omega') \big]
\nonumber\\
&\times&
\Im m\big[ G^{({\rm R})}_{\mu''}({\bm k}'_+-{\bm q}',\varepsilon'+\omega -\omega') \big]
\nonumber\\
&\times&
{\cal D}_{\lambda'\lambda''}({\bm k}_+,{\bm k}_+-{\bm q}')
{\cal D}_{\lambda''\lambda}({\bm k}_+-{\bm q}',{\bm k}_-)
\nonumber\\
&\times&
{\cal D}_{\mu\mu''}({\bm k}'_-,{\bm k}'_+-{\bm q}')
{\cal D}_{\mu''\mu'}({\bm k}'_+-{\bm q}',{\bm k}'_+)
~.
\end{eqnarray}
We can now take the limit $v_{\rm F} q\ll \omega \ll \varepsilon_{\rm F}$, and we get
\begin{eqnarray} \label{eq:SM_W_2_A-R_2}
&& \!\!\!\!\!\!\!\!
W^{(2)}_{\lambda\lambda'\mu\mu'} (\varepsilon'_--\varepsilon) - W^{(2)}_{\lambda\lambda'\mu\mu'} (\varepsilon'_+-\varepsilon) =
4 N_{\rm f} \sum_{{\bm q}',\lambda'',\mu''} 
\nonumber\\
&\times&
\int \frac{d\omega'}{2\pi i} 
\big|W({\bm q}',\omega')\big|^2 \big[n_{\rm F}(\omega'-\varepsilon) - n_{\rm F}(\omega'-\varepsilon')\big]
\nonumber\\
&\times&
\Im m \big[ G^{({\rm R})}_{\lambda''}({\bm k}-{\bm q}',\varepsilon-\omega') \big]
\Im m \big[ G^{({\rm R})}_{\mu''}({\bm k}'-{\bm q}',\varepsilon' -\omega') \big]
\nonumber\\
&\times&
{\cal D}_{\lambda'\lambda''}({\bm k},{\bm k}-{\bm q}')
{\cal D}_{\lambda''\lambda}({\bm k}-{\bm q}',{\bm k})
{\cal D}_{\mu\mu''}({\bm k}',{\bm k}'-{\bm q}')
\nonumber\\
&\times&
{\cal D}_{\mu''\mu'}({\bm k}'-{\bm q}',{\bm k}')
~.
\end{eqnarray}

\subsubsection{Analytical continuation of Eq.~(\ref{eq:SM_Lambda_3})}
We define $h(i\varepsilon_{n'}, i\varepsilon_{n'} + i\omega_m, i\varepsilon_{n'}+i\varepsilon_{n} +i\omega_m)$ such that
\begin{eqnarray} \label{eq:SM_Lambda_3_f}
&& \!\!\!\!\!\!\!\!
\Lambda^{(3)} ({\bm k}_+, i\varepsilon_n + i\omega_m,{\bm k}_-,i\varepsilon_n)
\nonumber\\
&\equiv&
- k_{\rm B} T \sum_{\varepsilon_{n'}} h(i\varepsilon_{n'}, i\varepsilon_{n'} + i\omega_m, i\varepsilon_{n'}+i\varepsilon_{n} +i\omega_m)
\nonumber\\
&=&
\oint \frac{dz}{2\pi i} n_{\rm F}(z) h(z, z + i\omega_m, z+i\varepsilon_{n} +i\omega_m)
~.
\nonumber\\
\end{eqnarray}
Here and in what follows we suppress for brevity all the band and spatial indices, and we retain only the index $i=3$. To perform the analytical continuation we first transform the sum over the poles of $n_{\rm F}(z)$ in an integration over the branch cuts of $h(z, z + i\omega_m, z+i\varepsilon_{n} +i\omega_m)$. We then analytically continue the result, according to the prescription $i\omega_m\to \omega_+$, $i\varepsilon_n\to \varepsilon_-$, $i\varepsilon_n+i\omega_m \to \varepsilon_++\omega$. After some lengthy algebra we get
\begin{eqnarray} \label{eq:SM_Lambda_3_f_BC}
&& \!\!\!\!\!\!\!\!
\Lambda^{(3)} ({\bm k}_+, \varepsilon_+ + \omega,{\bm k}_-,\varepsilon_-) =
\int \frac{d\varepsilon'}{2\pi i} \Big\{
n_{\rm F}(\varepsilon')
\nonumber\\
&\times& 
\big[ h(\varepsilon'_+, \varepsilon'_+ + \omega, \varepsilon'_++\varepsilon +\omega) 
\nonumber\\
&-&
h(\varepsilon'_-, \varepsilon'_+ + \omega, \varepsilon'_++\varepsilon +\omega) \big]
\nonumber\\
&+&
n_{\rm F}(\varepsilon') \big[ h(\varepsilon'_--\omega, \varepsilon'_+, \varepsilon'_-+\varepsilon) 
\nonumber\\
&-&
h(\varepsilon'_--\omega, \varepsilon'_-, \varepsilon'_-+\varepsilon) \big]
\nonumber\\
&-&
n_{\rm B}(\varepsilon') \big[ h(\varepsilon'_- -\varepsilon -\omega, \varepsilon'_+ - \varepsilon, \varepsilon'_+) 
\nonumber\\
&-&
h(\varepsilon'_--\varepsilon -\omega, \varepsilon'_+ -\varepsilon, \varepsilon'_-) \big] \Big\}
~.
\end{eqnarray}
We now shift $\varepsilon'\to \varepsilon'+\omega$ in the third and fourth lines of Eq.~(\ref{eq:SM_Lambda_3_f_BC}), and  we take the limit $\omega\to 0$ in $n_{\rm F}(\varepsilon'+\omega)$. We note that $h(\varepsilon'_+, \varepsilon'_+ + \omega, \varepsilon'_++\varepsilon +\omega)$ and $h(\varepsilon'_--\omega, \varepsilon'_-, \varepsilon'_-+\varepsilon)$ have the poles on the same half of the complex plane, and can be neglected in the limit $\varepsilon_{\rm F} \tau_{\rm qp}^{\rm ee} \gg 1$. We then shift $\varepsilon'\to \varepsilon'+\varepsilon+\omega$ in the last two lines of Eq.~(\ref{eq:SM_Lambda_3_f_BC}), and we take the limit $\omega \to 0$ in $n_{\rm B}(\varepsilon'+\varepsilon+\omega)$. After these manipulations Eq.~(\ref{eq:SM_Lambda_3_f_BC}) becomes
\begin{eqnarray} \label{eq:SM_Lambda_3_f_final}
&& \!\!\!\!\!\!\!\!
\Lambda^{(3)} ({\bm k}_+, \varepsilon_++\omega,{\bm k}_-,\varepsilon_-) =
\int \frac{d\varepsilon'}{2\pi i}
\big[n_{\rm F}(\varepsilon') + n_{\rm B}(\varepsilon' + \varepsilon)\big] 
\nonumber\\
&\times&
\big[ h(\varepsilon'_-, \varepsilon'_++\omega, \varepsilon'_-+\varepsilon+\omega) 
\nonumber\\
&-&
h(\varepsilon'_-, \varepsilon'_++\omega, \varepsilon'_++\varepsilon+\omega) \big]
\nonumber\\
&=&
\sum_{{\bm k}'} \sum_{\mu,\mu'} \int \frac{d\varepsilon'}{2\pi i} \big[n_{\rm F}(\varepsilon') + n_{\rm B}(\varepsilon' + \varepsilon)\big]
\nonumber\\
&\times&
\Big[W^{(3)}_{\lambda\lambda'\mu\mu'}({\bm k}, {\bm k}', \varepsilon'_-+\varepsilon) - W^{(3)}_{\lambda\lambda'\mu\mu'}({\bm k}, {\bm k}', \varepsilon'_++\varepsilon) \Big]
\nonumber\\
&\times&
G^{({\rm R})}_{\mu'}({\bm k}'_+,\varepsilon'+\omega) G^{({\rm A})}_{\mu}({\bm k}'_-,\varepsilon')
\nonumber\\
&\times&
\Lambda_{\mu'\mu,\beta} ({\bm k}'_+, \varepsilon_++\omega,{\bm k}'_-,\varepsilon'_-)
~.
\end{eqnarray}
It only remains to determine $W^{(3)}_{\lambda\lambda'\mu\mu'} ({\bm k}'-{\bm k},\varepsilon'_\pm-\varepsilon)$, defined in Eq.~(\ref{eq:SM_W_3}). Eq.~(\ref{eq:SM_Lambda_3_f_final}) implies that we have to analytically continue the functions $W^{(3)}$ for $i\varepsilon_{n'} \to \varepsilon'_-$ and $i\varepsilon_{n'} + i\omega_m \to \varepsilon'_+ + \omega$.

\subsubsection{The analytical continuation of Eq.~(\ref{eq:SM_W_3})}
We now turn to the analytical continuation of Eq.~(\ref{eq:SM_W_3}) with the prescription $i\omega_m\to \omega_+$, $i\varepsilon_n\to \varepsilon_-$, $i\varepsilon_n+i\omega_m \to \varepsilon_++\omega$, $i\varepsilon_{n'} \to \varepsilon'_-$, and $i\varepsilon_{n'} + i\omega_m \to \varepsilon'_+ + \omega$. We define
\begin{eqnarray} \label{eq:SM_W_3_f}
&& \!\!\!\!\!\!\!\!
W^{(3)}_{\lambda\lambda'\mu\mu'} ({\bm k}',{\bm k},i\varepsilon_{n'}+i\varepsilon_n+i\omega_m) \equiv
\oint \frac{dz}{2\pi i} n_{\rm B}(z)
\nonumber\\
&\times&
w_3(z,z-i\omega_m,i\varepsilon_n +z,i\varepsilon_{n'}+i\omega_m -z)
~.
\end{eqnarray}
Integrating over the branch cuts of $w_3(z,z-i\omega_m,i\varepsilon_n +z,i\varepsilon_{n'}+i\omega_m -z)$, and performing the analytical continuations according to the prescriptions stated before, we get
\begin{eqnarray} \label{eq:SM_W_3_f_BC}
&& \!\!\!\!\!\!\!\!
W^{(3)}_{\lambda\lambda'\mu\mu'} ({\bm k}',{\bm k},\varepsilon'_\pm+\varepsilon+\omega) =
\int \frac{d\omega'}{2\pi i} \Big\{
n_{\rm B}(\omega') 
\nonumber\\
&\times&
\big[w_3(\omega'_+,\omega'_--\omega,\omega'_- + \varepsilon, \varepsilon'+ \omega -\omega'_-)
\nonumber\\
&-&
w_3(\omega'_-,\omega'_--\omega,\omega'_- + \varepsilon, \varepsilon'+ \omega -\omega'_-)
\big]
\nonumber\\
&+&
n_{\rm B}(\omega') \big[w_3(\omega'_++\omega,\omega'_+,\omega'_++ \varepsilon+\omega, \varepsilon' -\omega'_+)
\nonumber\\
&-&
w_3(\omega'_++\omega,\omega'_-,\omega'_++ \varepsilon+\omega, \varepsilon' -\omega'_+)
\big]
\nonumber\\
&-&
n_{\rm F}(\omega')
\nonumber\\
&\times&
\big[w_3(\omega'_+-\varepsilon,\omega'_--\varepsilon-\omega,\omega'_+, \varepsilon'+\varepsilon+\omega -\omega'_\mp)
\nonumber\\
&-&
w_3(\omega'_+-\varepsilon,\omega'_--\varepsilon-\omega,\omega'_-, \varepsilon'+\varepsilon+\omega -\omega'_\mp)
\big]
\nonumber\\
&-&
n_{\rm F}(\omega')
\nonumber\\
&\times&
\big[w_3(\omega'_++\varepsilon'+\omega,\omega'_-+\varepsilon',\omega'_\pm + \varepsilon+\varepsilon'+\omega,  -\omega'_+)
\nonumber\\
&-&
w_3(\omega'_++\varepsilon'+\omega,\omega'_-+\varepsilon',\omega'_\pm + \varepsilon+\varepsilon'+\omega,  -\omega'_-)
\big]
\Big\}
~.
\nonumber\\
\end{eqnarray}
Note that the terms on the r.h.s. of Eq.~(\ref{eq:SM_W_3_f_BC}) proportional to $n_{\rm B}(\omega')$ are identical in both $W^{(3)}_{\lambda\lambda'\mu\mu'} ({\bm k}',{\bm k},\varepsilon'_\pm-\varepsilon)$ and thus vanish when the difference is taken. We will thus neglect these terms in what follows. With this choice, Eq.~(\ref{eq:SM_W_3_f_BC}) reduces to
\begin{eqnarray} \label{eq:SM_W_3_f_final}
&& \!\!\!\!\!\!\!\!
W^{(3)}_{\lambda\lambda'\mu\mu'} ({\bm k}',{\bm k},\varepsilon'_\pm+\varepsilon+\omega) =
-\int \frac{d\omega'}{2\pi i} \Big\{
n_{\rm F}(\omega'+\varepsilon)
\nonumber\\
&\times&
\big[w_3(\omega'_+,\omega'_--\omega,\omega'_++\varepsilon, \varepsilon'+\omega-\omega'_\mp)
\nonumber\\
&-&
w_3(\omega'_+,\omega'_--\omega,\omega'_-+\varepsilon, \varepsilon'+\omega-\omega'_\mp)
\big]
\nonumber\\
\!\!\! &+& \!\!\!
n_{\rm F}(\omega'-\varepsilon'-\omega) \big[w_3(\omega'_+,\omega'_--\omega,\omega'_\pm + \varepsilon,  \varepsilon'+\omega-\omega'_+)
\nonumber\\
&-&
w_3(\omega'_+,\omega'_--\omega,\omega'_\pm + \varepsilon,  \varepsilon'+\omega-\omega'_-)
\big]
\Big\}
~.
\nonumber\\
\end{eqnarray}
Finally, substituting Eq.~(\ref{eq:SM_W_3}) into Eq.~(\ref{eq:SM_W_3_f_final}) we get
\begin{eqnarray} \label{eq:SM_W_3_A-R}
&& \!\!\!\!\!\!\!\!
W^{(3)}_{\lambda\lambda'\mu\mu'} (\varepsilon'_-+\varepsilon+\omega) - W^{(3)}_{\lambda\lambda'\mu\mu'} (\varepsilon'_++\varepsilon+\omega) =
\int \frac{d\omega'}{2\pi i} 
\nonumber\\
&\times&
\big[n_{\rm F}(\omega'+\varepsilon) - n_{\rm F}(\omega'-\varepsilon'-\omega)\big]
\nonumber\\
&\times&
\big[
w_3(\omega'_+,\omega'_--\omega,\omega'_++\varepsilon, \varepsilon'+\omega-\omega'_-)
\nonumber\\
&-&
w_3(\omega'_+,\omega'_-,\omega'_-+\varepsilon, \varepsilon'+\omega-\omega'_-)
\nonumber\\
&-&
w_3(\omega'_+,\omega'_--\omega,\omega'_++\varepsilon, \varepsilon'+\omega-\omega'_+)
\nonumber\\
&+&
w_3(\omega'_+,\omega'_--\omega,\omega'_-+\varepsilon, \varepsilon'+\omega-\omega'_+)
\big]
\nonumber\\
&=&
-4 N_{\rm f} \sum_{{\bm q}'} \sum_{\lambda'',\mu''} \int \frac{d\omega'}{2\pi i} W({\bm q}',\omega'_+) W({\bm q}',\omega'_--\omega)
\nonumber\\
&\times&
\big[n_{\rm F}(\omega'+\varepsilon) - n_{\rm F}(\omega'-\varepsilon'-\omega)\big]
\nonumber\\
&\times&
\Im m \big[ G^{({\rm R})}_{\lambda''}({\bm k}_- +{\bm q}',\varepsilon+\omega') \big]
\nonumber\\
&\times&
\Im m \big[ G^{({\rm R})}_{\mu''}({\bm k}'_+ -{\bm q}',\varepsilon' +\omega- \omega') \big]
\nonumber\\
&\times&
{\cal D}_{\lambda\lambda''}({\bm k}_-,{\bm k}_-+{\bm q}')
{\cal D}_{\lambda''\lambda'}({\bm k}_-+{\bm q}',{\bm k}_+)
\nonumber\\
&\times&
{\cal D}_{\mu\mu''}({\bm k}'_-,{\bm k}'_+-{\bm q}')
{\cal D}_{\mu''\mu'}({\bm k}'_+-{\bm q}',{\bm k}'_+)
~.
\end{eqnarray}
Taking the limit $v_{\rm F} q \ll \omega\ll \varepsilon_{\rm F}$, Eq.~(\ref{eq:SM_W_3_A-R}) becomes
\begin{eqnarray}\label{eq:SM_W_3_A-R_2}
&& \!\!\!\!\!\!\!\!
W^{(3)}_{\lambda\lambda'\mu\mu'} (\varepsilon'_-+\varepsilon) - W^{(3)}_{\lambda\lambda'\mu\mu'} (\varepsilon'_++\varepsilon) =
-4 N_{\rm f} \sum_{{\bm q}',\lambda'',\mu''}
\nonumber\\
&\times&
\int \frac{d\omega'}{2\pi i} \big| W({\bm q}',\omega')\big|^2
\big[n_{\rm F}(\omega'+\varepsilon) - n_{\rm F}(\omega'-\varepsilon')\big]
\nonumber\\
&\times&
\Im m \big[ G^{({\rm R})}_{\lambda''} ({\bm k} +{\bm q}',\varepsilon+\omega') \big]
\Im m \big[ G^{({\rm R})}_{\mu''} ({\bm k}' -{\bm q}',\varepsilon' - \omega') \big]
\nonumber\\
&\times&
{\cal D}_{\lambda\lambda''}({\bm k},{\bm k}+{\bm q}')
{\cal D}_{\lambda''\lambda'}({\bm k}+{\bm q}',{\bm k})
{\cal D}_{\mu\mu''}({\bm k}',{\bm k}'-{\bm q}')
\nonumber\\
&\times&
{\cal D}_{\mu''\mu'}({\bm k}'-{\bm q}',{\bm k}')
~.
\end{eqnarray}

\subsection{The solution of the Bethe-Salpeter equation}
After the analytical continuation to real frequencies, and retaining only the dominant contribution in the limit of $v_{\rm F} q \ll \omega, (\tau_{\rm qp}^{\rm ee})^{-1} \ll \varepsilon_{\rm F}$,  the Bethe-Salpeter equation~(\ref{eq:SM_Lambda_def}) becomes
\begin{eqnarray} \label{eq:SM_BS_analytical}
\Lambda_{\lambda'\lambda,\beta} ({\bm k}, \varepsilon_+,{\bm k},\varepsilon_-) &=& \Lambda^{(0)}_{\lambda'\lambda,\beta} ({\bm k},\varepsilon_+,{\bm k},\varepsilon_-)
\nonumber\\
&+&
\sum_{i=1}^{3} \Lambda^{(i)}_{\lambda'\lambda,\beta} ({\bm k}, \varepsilon_+,{\bm k},\varepsilon_-)
~,
\nonumber\\
\end{eqnarray}
where $\big\{ \Lambda^{(i)}_{\lambda'\lambda,\beta} ({\bm k}, \varepsilon_+,{\bm k},\varepsilon_-),~ i=1,\ldots,3 \big\}$ are given in Eqs.~(\ref{eq:SM_Lambda_1_2_f_final}) and~(\ref{eq:SM_Lambda_3_f_final}), with the potentials $\big\{ W^{(i)}_{\lambda\lambda'\mu\mu'},~ i=1,\ldots,3 \big\}$ defined in Eqs.~(\ref{eq:SM_W_1_A-R_2}),~(\ref{eq:SM_W_2_A-R_2}) and~(\ref{eq:SM_W_3_A-R_2}). Putting everything together we find
\begin{eqnarray} \label{eq:SM_T_Lambda_1_f_final}
&& \!\!\!\!\!\!\!\!
\Lambda^{(1)}_{\lambda'\lambda,\beta} ({\bm k}, \varepsilon_+,{\bm k},\varepsilon_-) =
4 N_{\rm f} \sum_{{\bm k}',{\bm q}'}\sum_{\mu,\mu'} \sum_{\lambda'',\mu''}
\nonumber\\
&\times&
\int \frac{d\varepsilon'}{2\pi i} \int \frac{d\omega'}{2\pi i} 
|W({\bm k}-{\bm k}',\varepsilon'-\varepsilon)|^2
\nonumber\\
&\times&
\big[n_{\rm F}(\varepsilon') + n_{\rm B}(\varepsilon'-\varepsilon)\big]
\big[n_{\rm F} (\omega'+\varepsilon') - n_{\rm F} (\omega'+\varepsilon)\big] 
\nonumber\\
&\times&
\Im m \big[G^{({\rm R})}_{\lambda''}({\bm q}'-{\bm k},\omega'+\varepsilon)\big]
{\cal D}_{\mu\lambda}({\bm k}',{\bm k})
\nonumber\\
&\times&
\Im m \big[G^{({\rm R})}_{\mu''}({\bm q'}-{\bm k}', \omega'+\varepsilon')\big]
{\cal D}_{\lambda'\mu'}({\bm k},{\bm k}') 
\nonumber\\
&\times&
{\cal D}_{\lambda''\mu''}({\bm q}'-{\bm k},{\bm q}'-{\bm k}')
{\cal D}_{\mu''\lambda''}({\bm q}'-{\bm k}',{\bm q}'-{\bm k}) 
\nonumber\\
&\times&
G^{({\rm R})}_{\mu'}({\bm k}',\varepsilon'+\omega) G^{({\rm A})}_{\mu}({\bm k}',\varepsilon')
\Lambda_{\mu'\mu,\beta} ({\bm k}', \varepsilon'_+, {\bm k}',\varepsilon'_-)
\nonumber\\
&\equiv&
\int d\varepsilon'
\int d\omega'
\big[n_{\rm F}(\varepsilon') + n_{\rm B}(\varepsilon'-\varepsilon)\big]
\nonumber\\
&\times&
\big[n_{\rm F} (\omega'+\varepsilon') - n_{\rm F} (\omega'+\varepsilon)\big] 
f_1 (\varepsilon,\varepsilon',\omega')
~,
\end{eqnarray}
and
\begin{eqnarray} \label{eq:SM_T_Lambda_2_f_final}
&& \!\!\!\!\!\!\!\!
\Lambda^{(2)}_{\lambda'\lambda,\beta} ({\bm k}, \varepsilon_+,{\bm k},\varepsilon_-) =
4 N_{\rm f}\sum_{{\bm k}',{\bm q}'} \sum_{\mu,\mu'} \sum_{\lambda'',\mu''} 
\nonumber\\
&\times&
\int \frac{d\varepsilon'}{2\pi i} \int \frac{d\omega'}{2\pi i}
|W({\bm q}',\omega')|^2
\nonumber\\
&\times&
\big[n_{\rm F}(\varepsilon') + n_{\rm B}(\varepsilon'-\varepsilon)\big]
\big[n_{\rm F}(\omega'-\varepsilon) - n_{\rm F}(\omega'-\varepsilon')\big]
\nonumber\\
&\times&
\Im m \big[ G^{({\rm R})}_{\lambda''}({\bm k}-{\bm q}',\varepsilon-\omega') \big]
{\cal D}_{\lambda'\lambda''}({\bm k},{\bm k}-{\bm q}')
\nonumber\\
&\times&
\Im m \big[ G^{({\rm R})}_{\mu''}({\bm k}'-{\bm q}',\varepsilon' -\omega') \big]
{\cal D}_{\lambda''\lambda}({\bm k}-{\bm q}',{\bm k})
\nonumber\\
&\times&
{\cal D}_{\mu\mu''}({\bm k}',{\bm k}'-{\bm q}')
{\cal D}_{\mu''\mu'}({\bm k}'-{\bm q}',{\bm k}')
\nonumber\\
&\times&
G^{({\rm R})}_{\mu'}({\bm k}',\varepsilon'+\omega) G^{({\rm A})}_{\mu}({\bm k}',\varepsilon')
\Lambda_{\mu'\mu,\beta} ({\bm k}', \varepsilon'_+, {\bm k}',\varepsilon'_-)
\nonumber\\
&\equiv&
\int d\varepsilon'
\int d\omega'
\big[n_{\rm F}(\varepsilon') + n_{\rm B}(\varepsilon'-\varepsilon)\big]
\nonumber\\
&\times&
\big[n_{\rm F}(\omega'-\varepsilon) - n_{\rm F}(\omega'-\varepsilon')\big]
f_2 (\varepsilon,\varepsilon',\omega')
~,
\end{eqnarray}
and finally
\begin{eqnarray} \label{eq:SM_T_Lambda_3_f_final}
&& \!\!\!\!\!\!\!\!
\Lambda^{(3)}_{\lambda'\lambda,\beta} ({\bm k}, \varepsilon_+,{\bm k},\varepsilon_-) =
-4 N_{\rm f} \sum_{{\bm k}',{\bm q}'} \sum_{\mu,\mu'} \sum_{\lambda'',\mu''}
\nonumber\\
&\times&
\int \frac{d\varepsilon'}{2\pi i} \int \frac{d\omega'}{2\pi i} 
|W({\bm q}',\omega')|^2
\nonumber\\
&\times&
\big[n_{\rm F}(\varepsilon') + n_{\rm B}(\varepsilon' + \varepsilon)\big]
\big[n_{\rm F}(\omega'+\varepsilon) - n_{\rm F}(\omega'-\varepsilon')\big]
\nonumber\\
&\times&
\Im m\big[ G^{({\rm R})}_{\lambda''}({\bm k} +{\bm q}',\varepsilon+\omega') \big]
{\cal D}_{\lambda\lambda''}({\bm k},{\bm k}+{\bm q}')
\nonumber\\
&\times&
\Im m\big[ G^{({\rm R})}_{\mu''}({\bm k}' -{\bm q}',\varepsilon'- \omega') \big]
{\cal D}_{\lambda''\lambda'}({\bm k}+{\bm q}',{\bm k})
\nonumber\\
&\times&
{\cal D}_{\mu\mu''}({\bm k}',{\bm k}'-{\bm q}')
{\cal D}_{\mu''\mu'}({\bm k}'-{\bm q}',{\bm k}')
\nonumber\\
&\times&
G^{({\rm R})}_{\mu'} ({\bm k}',\varepsilon'+\omega)
G^{({\rm A})}_{\mu} ({\bm k}',\varepsilon')
\Lambda_{\mu'\mu,\beta} ({\bm k}', \varepsilon'_+,{\bm k}',\varepsilon'_-)
\nonumber\\
&\equiv&
\int d\varepsilon'
\int d\omega'
\big[n_{\rm F}(\varepsilon') + n_{\rm B}(\varepsilon' + \varepsilon)\big]
\nonumber\\
&\times&
\big[n_{\rm F}(\omega'+\varepsilon) - n_{\rm F}(\omega'-\varepsilon')\big]
f_3 (\varepsilon,\varepsilon',\omega')
~.
\end{eqnarray}

In the above equations we have defined
\begin{eqnarray} \label{eq:SM_T_Lambda_1_f_1_def}
&& \!\!\!\!\!\!\!\!
f_1 (\varepsilon,\varepsilon',\omega')
 =
- \frac{N_{\rm f}}{\pi^2}  \sum_{{\bm k}',{\bm q}'}\sum_{\mu,\mu'} \sum_{\lambda'',\mu''}
|W({\bm k}-{\bm k}',\varepsilon'-\varepsilon)|^2
\nonumber\\
&\times&
\Im m \big[G^{({\rm R})}_{\lambda''}({\bm q}'-{\bm k},\omega'+\varepsilon)\big]
{\cal D}_{\mu\lambda}({\bm k}',{\bm k})
\nonumber\\
&\times&
\Im m \big[G^{({\rm R})}_{\mu''}({\bm q'}-{\bm k}', \omega'+\varepsilon')\big]
{\cal D}_{\lambda'\mu'}({\bm k},{\bm k}') 
\nonumber\\
&\times&
{\cal D}_{\lambda''\mu''}({\bm q}'-{\bm k},{\bm q}'-{\bm k}')
{\cal D}_{\mu''\lambda''}({\bm q}'-{\bm k}',{\bm q}'-{\bm k}) 
\nonumber\\
&\times&
G^{({\rm R})}_{\mu'}({\bm k}',\varepsilon'+\omega) G^{({\rm A})}_{\mu}({\bm k}',\varepsilon')
\Lambda_{\mu'\mu,\beta} ({\bm k}', \varepsilon'_+, {\bm k}',\varepsilon'_-)
~,
\nonumber\\
\end{eqnarray}
and
\begin{eqnarray} \label{eq:SM_T_Lambda_2_f_2_def}
&& \!\!\!\!\!\!\!\!
f_2(\varepsilon,\varepsilon',\omega') =
-\frac{N_{\rm f}}{\pi^2} \sum_{{\bm k}',{\bm q}'} \sum_{\mu,\mu'} \sum_{\lambda'',\mu''} 
|W({\bm q}',\omega')|^2
\nonumber\\
&\times&
\Im m \big[ G^{({\rm R})}_{\lambda''}({\bm k}-{\bm q}',\varepsilon-\omega') \big]
{\cal D}_{\lambda'\lambda''}({\bm k},{\bm k}-{\bm q}')
\nonumber\\
&\times&
\Im m \big[ G^{({\rm R})}_{\mu''}({\bm k}'-{\bm q}',\varepsilon' -\omega') \big]
{\cal D}_{\lambda''\lambda}({\bm k}-{\bm q}',{\bm k})
\nonumber\\
&\times&
{\cal D}_{\mu\mu''}({\bm k}',{\bm k}'-{\bm q}')
{\cal D}_{\mu''\mu'}({\bm k}'-{\bm q}',{\bm k}')
\nonumber\\
&\times&
G^{({\rm R})}_{\mu'}({\bm k}',\varepsilon'+\omega) G^{({\rm A})}_{\mu}({\bm k}',\varepsilon')
\Lambda_{\mu'\mu,\beta} ({\bm k}', \varepsilon'_+, {\bm k}',\varepsilon'_-)
~,
\nonumber\\
\end{eqnarray}
and finally
\begin{eqnarray} \label{eq:SM_T_Lambda_3_f_3_def}
&& \!\!\!\!\!\!\!\!
f_3 (\varepsilon,\varepsilon',\omega') =
\frac{N_{\rm f}}{\pi^2} \sum_{{\bm k}',{\bm q}'} \sum_{\mu,\mu'} \sum_{\lambda'',\mu''}
|W({\bm q}',\omega')|^2
\nonumber\\
&\times&
\Im m\big[ G^{({\rm R})}_{\lambda''}({\bm k} +{\bm q}',\varepsilon+\omega') \big]
{\cal D}_{\lambda\lambda''}({\bm k},{\bm k}+{\bm q}')
\nonumber\\
&\times&
\Im m\big[ G^{({\rm R})}_{\mu''}({\bm k}' -{\bm q}',\varepsilon'- \omega') \big]
{\cal D}_{\lambda''\lambda'}({\bm k}+{\bm q}',{\bm k})
\nonumber\\
&\times&
{\cal D}_{\mu\mu''}({\bm k}',{\bm k}'-{\bm q}')
{\cal D}_{\mu''\mu'}({\bm k}'-{\bm q}',{\bm k}')
\nonumber\\
&\times&
G^{({\rm R})}_{\mu'} ({\bm k}',\varepsilon'+\omega)
G^{({\rm A})}_{\mu} ({\bm k}',\varepsilon')
\Lambda_{\mu'\mu,\beta} ({\bm k}', \varepsilon'_+,{\bm k}',\varepsilon'_-)
~.
\nonumber\\
\end{eqnarray}

We now expand Eqs.~(\ref{eq:SM_T_Lambda_1_f_final})-(\ref{eq:SM_T_Lambda_3_f_final}) for small $\varepsilon$, and we consider the term of this expansion that is linear in $\varepsilon$. The term of ${\cal O}(\varepsilon^0)$ does not contribute to the thermal conductivity. In expanding Eqs.~(\ref{eq:SM_T_Lambda_1_f_final})-(\ref{eq:SM_T_Lambda_3_f_final}) we have two possibilities. We can either expand the square brackets which contain the Bose and Fermi distributions, and replace $f_i(\varepsilon,\varepsilon',\omega') \to f_i(0,\varepsilon',\omega')$, or expand the functions $f_i(\varepsilon,\varepsilon',\omega')$. Let us consider the latter case.
We thus expand $f_i(\varepsilon,\varepsilon',\omega')$ to ${\cal O}(\varepsilon)$, and we set $\varepsilon=0$ in all the Fermi and Bose distributions. Each integrand contains the function $\big[n_{\rm F}(\varepsilon') + n_{\rm B}(\varepsilon')\big]$, which is peaked at $\varepsilon'=0$. We thus consider the remaining functions in the limit of small $\varepsilon'$. In this limit the difference of the two Fermi functions in Eqs.~(\ref{eq:SM_T_Lambda_1_f_final})-(\ref{eq:SM_T_Lambda_3_f_final}) is linear in $\varepsilon'$. Thus, when we expand $f_i(\varepsilon,\varepsilon',\omega')$ to ${\cal O}(\varepsilon)$ we always get contributions of the type
\begin{equation} \label{eq:SM_f_i_derivative_T4}
\int d\varepsilon'
\big[n_{\rm F}(\varepsilon') + n_{\rm B}(\varepsilon')\big] \varepsilon'
\left.\frac{\partial f_i (\varepsilon,\varepsilon',\omega')}{\partial \varepsilon} \right|_{\varepsilon=0}
= {\cal O}(T^3)
~.
\end{equation}
The equality holds because $f_i (\varepsilon,\varepsilon',\omega') \propto \varepsilon'$ [since it contains the function $\Lambda_{\mu'\mu,\beta} ({\bm k}'_+, \varepsilon_+,{\bm k}'_-,\varepsilon'_-) \propto \varepsilon'$]. This in turn implies that, since the integral on the left-hand side of Eq.~(\ref{eq:SM_f_i_derivative_T4}) is proportional to $\varepsilon'^2$, it naturally scales as $T^3$ as can be easily seen by introducing the dimensionless variable ${\bar \varepsilon}' = \varepsilon'/(k_{\rm B} T)$. On the contrary, when we expand to ${\cal O}(\varepsilon)$ the square brackets containing the Bose and Fermi distributions on the right-hand side of Eqs.~(\ref{eq:SM_T_Lambda_1_f_final})-(\ref{eq:SM_T_Lambda_3_f_final}), we obtain terms which scale with $T^2$. In the low-temperature limit we are thus naturally led to neglect contributions like that of Eq.~(\ref{eq:SM_f_i_derivative_T4}) and to replace $f_i(\varepsilon,\varepsilon',\omega') \to f_i(0,\varepsilon',\omega')$ in Eqs.~(\ref{eq:SM_T_Lambda_1_f_final})-(\ref{eq:SM_T_Lambda_3_f_final}). Moreover, when $\omega \to 0$ we can write
\begin{equation}
f_i (\varepsilon,\varepsilon',\omega') = \varepsilon' \left. \frac{\partial f_i (0,\varepsilon',0)}{\partial \varepsilon'} \right|_{\varepsilon'=0}
~,
\end{equation}
owing to the fact that the combination of Fermi and Bose distributions is strongly peaked at $\varepsilon \sim \varepsilon' \sim \omega' \sim 0$. With this approximation, Eqs.~(\ref{eq:SM_T_Lambda_1_f_final})-(\ref{eq:SM_T_Lambda_3_f_final}) read
\begin{eqnarray} \label{eq:SM_T_Lambda_1_f_final_2}
&& \!\!\!\!\!\!\!\!
\Lambda^{(1)}_{++,\beta} ({\bm k}, \varepsilon_+,{\bm k},\varepsilon_-) =
-\frac{8 i N_{\rm f}}{\omega + i/\tau_{\rm qp}^{\rm ee}}
\int \frac{d\varepsilon'}{2\pi i} \int \frac{d\omega'}{2\pi i} \varepsilon' 
\nonumber\\
&\times&
\big[n_{\rm F}(\varepsilon') + n_{\rm B}(\varepsilon'-\varepsilon)\big]
\big[n_{\rm F} (\omega'+\varepsilon') - n_{\rm F} (\omega'+\varepsilon)\big] 
\nonumber\\
&\times&
\sum_{{\bm k}',{\bm q}'}
|W({\bm k}-{\bm k}',0)|^2
\Im m\big[ G^{({\rm R})}_{+}({\bm k}', 0) \big]
\nonumber\\
&\times&
\Im m\big[ G^{({\rm R})}_{+}({\bm q}'-{\bm k},0)\big]
\Im m\big[ G^{({\rm R})}_{+}({\bm q'}-{\bm k}',0)\big]
\nonumber\\
&\times&
{\cal D}_{++}({\bm q}'-{\bm k},{\bm q}'-{\bm k}')
{\cal D}_{++}({\bm q}'-{\bm k}',{\bm q}'-{\bm k}) 
\nonumber\\
&\times&
{\cal D}_{++}({\bm k}',{\bm k})
{\cal D}_{++}({\bm k},{\bm k}') 
\partial_{\varepsilon''} \Lambda_{++,\beta} ({\bm k}', \varepsilon''_+, {\bm k}',\varepsilon''_-)\Big|_{\varepsilon''=0}
~,
\nonumber\\
\end{eqnarray}
and
\begin{eqnarray} \label{eq:SM_T_Lambda_2_f_final_2}
&& \!\!\!\!\!\!\!\!
\Lambda^{(2)}_{++,\beta} ({\bm k}, \varepsilon_+,{\bm k},\varepsilon_-) =
-\frac{8 i N_{\rm f}}{\omega + i/\tau_{\rm qp}^{\rm ee}}
\int \frac{d\varepsilon'}{2\pi i} \int \frac{d\omega'}{2\pi i}\varepsilon' 
\nonumber\\
&\times&
\big[n_{\rm F}(\varepsilon') + n_{\rm B}(\varepsilon'-\varepsilon)\big]
\big[n_{\rm F}(\omega'+\varepsilon')- n_{\rm F}(\omega'+\varepsilon)\big]
\nonumber\\
&\times&
\sum_{{\bm k}',{\bm q}'}
|W({\bm k}-{\bm k}',0)|^2
\Im m \big[ G^{({\rm R})}_{+}({\bm k}',0) \big]
\nonumber\\
&\times&
\Im m \big[ G^{({\rm R})}_{+}({\bm k}'-{\bm q}',0) \big]
\Im m \big[G^{({\rm R})}_{+}({\bm k}-{\bm q}',0) \big]
\nonumber\\
&\times&
{\cal D}_{++}({\bm k}'-{\bm q}',{\bm k}-{\bm q}')
{\cal D}_{++}({\bm k}-{\bm q}',{\bm k}'-{\bm q}')
\nonumber\\
&\times&
{\cal D}_{++}({\bm k},{\bm k}')
{\cal D}_{++}({\bm k}',{\bm k})
\nonumber\\
&\times&
\partial_{\varepsilon''} \Lambda_{++,\beta} ({\bm k}-{\bm q}', \varepsilon''_+, {\bm k}-{\bm q}',\varepsilon''_-) \Big|_{\varepsilon''=0}
~,
\end{eqnarray}
and finally
\begin{eqnarray} \label{eq:SM_T_Lambda_3_f_final_2}
&& \!\!\!\!\!\!\!\!
\Lambda^{(3)}_{++,\beta} ({\bm k}, \varepsilon_+,{\bm k},\varepsilon_-) =
- \frac{8 i N_{\rm f}}{\omega + i/\tau_{\rm qp}^{\rm ee}}
\int \frac{d\varepsilon'}{2\pi i} \int \frac{d\omega'}{2\pi i} \varepsilon' 
\nonumber\\
&\times&
\big[n_{\rm F}(\varepsilon') + n_{\rm B}(\varepsilon' - \varepsilon)\big]
\big[n_{\rm F}(\omega'+\varepsilon') - n_{\rm F}(\omega'+\varepsilon)\big]
\nonumber\\
&\times&
\sum_{{\bm k}',{\bm q}'}
|W({\bm k}'-{\bm k},0)|^2
\Im m\big[ G^{({\rm R})}_{+}({\bm k}',0) \big]
\nonumber\\
&\times&
\Im m\big[ G^{({\rm R})}_{+}({\bm k}-{\bm q}',0) \big]
\Im m \big[G^{({\rm R})}_{+} ({\bm k}'-{\bm q}',0)\big]
\nonumber\\
&\times&
{\cal D}_{++}({\bm k}'-{\bm q}',{\bm k}-{\bm q}')
{\cal D}_{++}({\bm k}-{\bm q}',{\bm k}'-{\bm q}')
\nonumber\\
&\times&
{\cal D}_{++}({\bm k},{\bm k}')
{\cal D}_{++}({\bm k}',{\bm k})
\nonumber\\
&\times&
\partial_{\varepsilon''}\Lambda_{++,\beta} ({\bm k}'-{\bm q}', \varepsilon''_+,{\bm k}'-{\bm q}',\varepsilon''_-)\Big|_{\varepsilon''=0}
~.
\end{eqnarray}
In these equations we used the crucial Fermi liquid identity
\begin{equation} \label{eq:SM_GR_GA_product}
G^{({\rm A})}_\lambda({\bm k}, \varepsilon) G^{({\rm R})}_{\lambda'}({\bm k}, \varepsilon+\omega) \to
-\frac{2 i \delta_{\lambda,+} \delta_{\lambda',+} }{\omega + i/\tau_{\rm qp}^{\rm ee}}
\Im m\big[ G^{({\rm R})}_{+}({\bm k}, 0) \big]\,,
\end{equation}
and the fact that the momenta of the retarded Green's function are all bounded to the Fermi surface.   Since the system we are considering is n-doped, the band index carried by the Green's function must refer to the conduction band. Finally, in Eq.~(\ref{eq:SM_T_Lambda_2_f_final_2}) we shifted ${\bm k}' \to {\bm k}-{\bm q}'$, ${\bm q}' \to {\bm k}-{\bm k}'$ and $\omega'\to -\omega'$, while in Eq.~(\ref{eq:SM_T_Lambda_3_f_final_2}) we replaced ${\bm k}' \to {\bm k}'-{\bm q}'$, ${\bm q}'\to{\bm k}'-{\bm k}$ and $\varepsilon' \to -\varepsilon'$.

Putting Eqs.~(\ref{eq:SM_T_Lambda_1_f_final_2})-(\ref{eq:SM_T_Lambda_3_f_final_2}) into Eq.~(\ref{eq:SM_BS_analytical}) we finally get
\begin{widetext}
\begin{eqnarray} \label{eq:SM_T_Bethe_Salpeter}
\Lambda_{++,\beta} ({\bm k}, \varepsilon_+,{\bm k},\varepsilon_-) &=& \Lambda^{(0)}_{++,\beta} ({\bm k},\varepsilon_+,{\bm k},\varepsilon_-)
- \frac{8 i N_{\rm f}}{\omega + i/\tau_{\rm qp}^{\rm ee}}
\int \frac{d\varepsilon'}{2\pi i} \int \frac{d\omega'}{2\pi i} \varepsilon' 
\big[n_{\rm F}(\varepsilon') + n_{\rm B}(\varepsilon' - \varepsilon)\big]
\nonumber\\
&\times&
\big[n_{\rm F}(\omega'+\varepsilon) - n_{\rm F}(\omega'+\varepsilon')\big]
\sum_{{\bm k}',{\bm q}'}
|W({\bm k}-{\bm k}',0)|^2
\Im m \Big[G^{({\rm R})}_{+}({\bm q}'-{\bm k},0)\Big]
\nonumber\\
&\times&
\Im m \Big[G^{({\rm R})}_{+}({\bm q'}-{\bm k}', 0)\Big]
\Im m \big[G^{({\rm R})}_{+}({\bm k}',0)\big]
{\cal D}_{++}({\bm k},{\bm k}') {\cal D}_{++}({\bm k}',{\bm k})
\nonumber\\
&\times&
{\cal D}_{++}({\bm q}'-{\bm k},{\bm q}'-{\bm k}') {\cal D}_{++}({\bm q}'-{\bm k}',{\bm q}'-{\bm k}) 
\Big[
\partial_{\varepsilon''}\Lambda_{++,\beta} ({\bm k}', \varepsilon''_+, {\bm k}',\varepsilon''_-)
\nonumber\\
&+&
\partial_{\varepsilon''} \Lambda_{++,\beta} ({\bm k}-{\bm q}', \varepsilon''_+, {\bm k}-{\bm q}',\varepsilon''_-) 
+
\partial_{\varepsilon''}\Lambda_{++,\beta} ({\bm k}'-{\bm q}', \varepsilon''_+,{\bm k}'-{\bm q}',\varepsilon''_-)
\Big]_{\varepsilon''=0}
~.
\end{eqnarray}
\end{widetext}
Eq.~(\ref{eq:SM_T_Bethe_Salpeter}) can be solved by the usual {\it ansatz}
\begin{equation} \label{eq:SM_T_vertex_ansatz}
\Lambda_{++,\beta} ({\bm k}, \varepsilon_+,{\bm k},\varepsilon_-) = \gamma(\omega) \Lambda^{(0)}_{++,\beta} ({\bm k},\varepsilon_+,{\bm k},\varepsilon_-)
~.
\end{equation}
Recalling that $\Lambda^{(0)}_{++,\beta} ({\bm k},\varepsilon_+,{\bm k},\varepsilon_-) = \varepsilon {\hat {\bm k}}_\beta$, Eq.~(\ref{eq:SM_T_Bethe_Salpeter}) becomes
\begin{eqnarray} \label{eq:SM_T_Bethe_Salpeter_2}
&& \!\!\!\!\!\!\!\!
\gamma(\omega) \varepsilon {\hat {\bm k}} = 
\varepsilon {\hat {\bm k}} 
- \gamma(\omega) \frac{8 i N_{\rm f}}{\omega + i/\tau_{\rm qp}^{\rm ee}}
\int \frac{d\varepsilon'}{2\pi i} \int \frac{d\omega'}{2\pi i} \varepsilon' 
\nonumber\\
&\times&
\big[n_{\rm F}(\varepsilon') + n_{\rm B}(\varepsilon' - \varepsilon)\big]
\big[n_{\rm F}(\omega'+\varepsilon) - n_{\rm F}(\omega'+\varepsilon')\big]
\nonumber\\
&\times&
\sum_{{\bm k}',{\bm q}'}
|W({\bm k}-{\bm k}',0)|^2
\Im m \big[G^{({\rm R})}_{+}({\bm k}',0)\big]
\nonumber\\
&\times&
\Im m \big[G^{({\rm R})}_{+}({\bm q}'-{\bm k},0)\big]
\Im m \big[G^{({\rm R})}_{+}({\bm q'}-{\bm k}', 0)\big]
\nonumber\\
&\times&
{\cal D}_{++}({\bm q}'-{\bm k},{\bm q}'-{\bm k}') {\cal D}_{++}({\bm q}'-{\bm k}',{\bm q}'-{\bm k}) 
\nonumber\\
&\times&
{\cal D}_{++}({\bm k},{\bm k}') {\cal D}_{++}({\bm k}',{\bm k})
\frac{ 2 {\bm k}' - 2 {\bm q}' + {\bm k} }{k_{\rm F}}
~.
\end{eqnarray}
Here we have used the fact that the wave vectors ${\bm k}-{\bm q}'$,  ${\bm k}'-{\bm q}'$ and ${\bm k}'$ on the right-hand side of Eq.~(\ref{eq:SM_T_Bethe_Salpeter_2}) are all bounded to the Fermi surface.  This allowed us to write, e.g.
\begin{eqnarray}
\frac{{\bm k}-{\bm q}'}{|{\bm k}-{\bm q}'|} = \frac{{\bm k}-{\bm q}'}{k_{\rm F}}
~,
\end{eqnarray}
and similarly for ${\bm k}' - {\bm q}'$ and ${\bm k}'$. Eq.~(\ref{eq:SM_T_Bethe_Salpeter_2}) should be plugged into Eq.~(\ref{eq:SM_chi_jj_final_omega_finite}), which in the limit $\omega \to 0$ reads
\begin{eqnarray} \label{eq:SM_chi_jj_final_omega0}
&& \!\!\!\!\!\!\!\!
\chi_{j^{({\rm Q})}_{\alpha} j^{({\rm Q})}_{\beta}} ({\bm q}={\bm 0}, \omega) \to \frac{2 i \omega N_{\rm f}}{\omega + i/\tau_{\rm qp}^{\rm ee}} \sum_{{\bm k}} \int \frac{d\varepsilon}{2\pi i} \frac{\partial n_{\rm F} (\varepsilon)}{\partial \varepsilon} \varepsilon
\nonumber\\
&\times&
\Im m \big[G^{({\rm R})}_{+}({\bm k}, 0)\big]
{\bm k}_\alpha
\Lambda_{++,\beta} ({\bm k}, \varepsilon_+,{\bm k},\varepsilon_-)
~.
\end{eqnarray}
We now observe that the angular integration picks only the component of $\Lambda_{++,\beta} ({\bm k}, \varepsilon_+,{\bm k},\varepsilon_-)$ parallel to ${\bm k}_\alpha$. Moreover, we can perform the frequency integration of Eqs.~(\ref{eq:SM_chi_jj_final_omega0}) and~(\ref{eq:SM_T_Bethe_Salpeter_2}) with the help of the following integrals
\begin{equation}
{\cal I}_1 \equiv \int d\varepsilon \left(-\frac{\partial n_{\rm F} (\varepsilon)}{\partial \varepsilon}\right) \varepsilon^2 = \frac{\pi^2}{3}(k_{\rm B} T)^2
~,
\end{equation}
and
\begin{eqnarray}
{\cal I}_2 &\equiv& \int d\varepsilon \int d\varepsilon' \int d\omega' \varepsilon \varepsilon' 
\left(-\frac{\partial n_{\rm F} (\varepsilon)}{\partial \varepsilon}\right)
\nonumber\\
&\times&
\big[n_{\rm F}(\varepsilon') + n_{\rm B}(\varepsilon' - \varepsilon)\big]
\big[n_{\rm F}(\omega'+\varepsilon) - n_{\rm F}(\omega'+\varepsilon')\big]
\nonumber\\
&=&
\frac{2\pi^4}{15} (k_{\rm B} T)^4
~.
\end{eqnarray}
Eq.~(\ref{eq:SM_T_Bethe_Salpeter_2}) can thus be rewritten as
\begin{eqnarray} \label{eq:SM_T_Bethe_Salpeter_3}
\gamma(\omega) &=& 
1 - \gamma(\omega) \frac{4 i N_{\rm f} (k_{\rm B} T)^2}{5(\omega + i/\tau_{\rm qp}^{\rm ee})} 
\sum_{{\bm k}',{\bm q}'}
|W({\bm k}-{\bm k}',0)|^2
\nonumber\\
&\times&
\Im m \big[G^{({\rm R})}_{+}({\bm k}',0)\big]
\Im m \big[G^{({\rm R})}_{+}({\bm q}'-{\bm k},0)\big]
\nonumber\\
&\times&
\Im m \big[G^{({\rm R})}_{+}({\bm q'}-{\bm k}', 0)\big]
{\cal D}_{++}({\bm k},{\bm k}') {\cal D}_{++}({\bm k}',{\bm k})
\nonumber\\
&\times&
{\cal D}_{++}({\bm q}'-{\bm k},{\bm q}'-{\bm k}') {\cal D}_{++}({\bm q}'-{\bm k}',{\bm q}'-{\bm k}) 
\nonumber\\
&\times&
\big[ 1 + 2 \cos(\varphi_{{\bm k}' - {\bm q}'} - \varphi_{\bm k}) \big]
~.
\end{eqnarray}
Defining
\begin{eqnarray} \label{eq:SM_thermal_transport_time}
\frac{1}{\tau_{\rm th}^{\rm ee}} &=& - \frac{32}{15} N_{\rm f}(k_{\rm B} T)^2
\sum_{{\bm k}',{\bm q}'}
|W({\bm k}-{\bm k}',0)|^2
\nonumber\\
&\times&
\Im m \big[G^{({\rm R})}_{+}({\bm q}'-{\bm k},0)\big]
\Im m \big[G^{({\rm R})}_{+}({\bm q'}-{\bm k}', 0)\big]
\nonumber\\
&\times&
\Im m \big[G^{({\rm R})}_{+}({\bm k}',0)\big]
{\cal D}_{++}({\bm k},{\bm k}') {\cal D}_{++}({\bm k}',{\bm k})
\nonumber\\
&\times&
{\cal D}_{++}({\bm q}'-{\bm k},{\bm q}'-{\bm k}') {\cal D}_{++}({\bm q}'-{\bm k}',{\bm q}'-{\bm k}) 
\nonumber\\
&\times&
\left\{ 1 - \frac{3}{4}\big[1 + \cos(\varphi_{{\bm k}' - {\bm q}'} - \varphi_{\bm k})\big] \right\}
~,
\end{eqnarray}
we get
\begin{equation} \label{eq:SM_T_Bethe_Salpeter_4}
\gamma(\omega) = 1 + \gamma(\omega) \frac{i/\tau_{\rm qp}^{\rm ee} - i/\tau_{\rm th}^{\rm ee}}{\omega+i/\tau_{\rm qp}^{\rm ee}}
~,
\end{equation}
which is solved by
\begin{equation} \label{eq:SM_T_Bethe_Salpeter_final}
\gamma(\omega) = \frac{\omega+i/\tau_{\rm qp}^{\rm ee}}{\omega + i/\tau_{\rm th}^{\rm ee}}
~.
\end{equation}
Note that the energy arguments of $W({\bm k}-{\bm k}',0)$, $\Im m \big[G^{({\rm R})}_{+}({\bm q'}-{\bm k}', 0)\big]$ and $\Im m \big[G^{({\rm R})}_{+}({\bm k}',0)\big]$ in Eq.~(\ref{eq:SM_thermal_transport_time}) should be proportional to the temperature $T$ rather than be exactly zero. This is accomplished with the same trick explained in Sect.~\ref{app:QP_lifetime}.

\section{The connection between the thermal transport time and the quasiparticle lifetime}
\label{app:thermal_time}
In this section we prove that the thermal transport time of Eq.~(\ref{eq:SM_thermal_transport_time}) is proportional to the quasiparticle lifetime of Eq.~(\ref{eq:SM_inverse_lifetime}). 

Let us consider Eq.~(\ref{eq:SM_thermal_transport_time}).  After performing the shifts ${\bm k}' \to {\bm k} - {\bm q}$ and ${\bm q}' \to -{\bm k}''+{\bm k}$ it reads
\begin{eqnarray} \label{eq:SM_thermal_transport_time_shift}
\frac{1}{\tau_{\rm th}^{\rm ee}} &=& - \frac{32}{15} N_{\rm f}(k_{\rm B} T)^2
\sum_{{\bm k}'',{\bm q}}
|W({\bm q},{\bar \varepsilon})|^2
\Im m \big[G^{({\rm R})}_{+}({\bm k}'',{\bar \varepsilon})\big]
\nonumber\\
&\times&
\Im m \big[G^{({\rm R})}_{+}({\bm k}''-{\bm q}, 0)\big]
\Im m \big[G^{({\rm R})}_{+}({\bm k} -{\bm q},{\bar \varepsilon})\big]
\nonumber\\
&\times&
{\cal D}_{++}({\bm k},{\bm k}-{\bm q}) {\cal D}_{++}({\bm k}-{\bm q},{\bm k})
\nonumber\\
&\times&
{\cal D}_{++}({\bm k}'',{\bm k}''-{\bm q}) {\cal D}_{++}({\bm k}''-{\bm q},{\bm k}'') 
\nonumber\\
&\times&
\left\{ 1 + \frac{3}{4}\big[1 + \cos(\varphi_{{\bm k}'' - {\bm q}} - \varphi_{\bm k})\big] \right\}
~,
\end{eqnarray}
which describes the relaxation of a thermal current due to the excitation of two particle-hole pairs, namely ${\bm k} \to {\bm k}-{\bm q}$ and ${\bm k}''-{\bm q} \to {\bm k}''$. 
The only difference between this expression and the expression~(\ref{eq:SM_inverse_lifetime}) for the quasiparticle lifetime, apart from the numerical factor,  is the presence of the factor $\big[1 + \cos(\varphi_{{\bm k}'' - {\bm q}} - \varphi_{\bm k})\big]$ in the integrand of  Eq.~(\ref{eq:SM_thermal_transport_time_shift}).  As we showed in Fig.~\ref{fig:SM_two}, the momentum conservation constrains ${\bm k}$ and ${\bm k}'' - {\bm q}$ to be diametrically opposite. This in turn implies that $\cos(\varphi_{{\bm k}'' - {\bm q}} - \varphi_{\bm k}) = -1$.  At last, the ratio between the thermal relaxation rate and the quasiparticle decay rate is seen to be 
\begin{equation}
\frac{1/\tau_{\rm th}^{\rm ee}}{1/\tau_{\rm qp}^{\rm ee}} = \frac{32/15}{4/3}=\frac{8}{5}
~.
\end{equation}
%

%%%%%%%%%%%%%%%%%%%
\begin{figure}[t]
\begin{center}
\begin{tabular}{c}
\includegraphics[width=0.99\columnwidth]{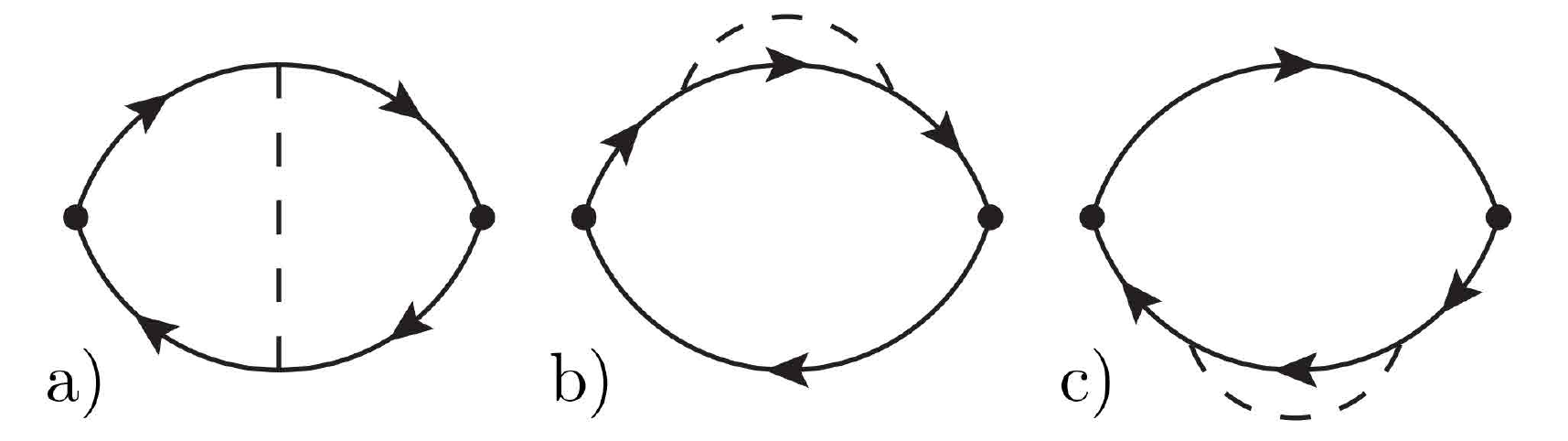}
\end{tabular}
\end{center}
\caption{
A diagrammatic representation of the first-order calculation performed in Sect.~\ref{app:renormalization_drude_weight}. Panel~a) shows the vertex correction to the energy-density linear response function, while panels~b) and~c) show the two time-reversal-related self-energy corrections. Solid lines represent bare Green's functions, while dashed lines stand for bare e-e interactions. The solid dots are bare energy vertices. The left vertex of all the three diagrams is always $\propto \xi_{{\bm k}_-}+\xi_{{\bm k}_+}$. The right vertex is instead $\propto \xi_{{\bm k}'_-}+\xi_{{\bm k}'_+}$ for the diagram in panel~a) and $\propto \xi_{{\bm k}_-}+\xi_{{\bm k}_+}$ for the diagrams in panels~b) and~c). Here ${\bm q}$ is the wavevectors of the external perturbation, while ${\bm k}$ and ${\bm k}'$ are internal particle momenta. Note that, since this is a first order calculation, the vertices are not renormalized by e-e interactions.
\label{fig:SM_five}}
\end{figure}
%%%%%%%%%%%%%%%%%%%

\section{The renormalization of the thermal Drude weight to the first order in the strength of e-e interactions}
\label{app:renormalization_drude_weight}
In this section we determine the first-order renormalization of the thermal Drude weight from the energy-density linear response function. We consider the three diagrams depicted in Fig.~\ref{fig:SM_five}a)-c). Analogous diagrams for the number-density response function were studied in Ref.~\onlinecite{Abedinpour_prb_2011} to calculate the renormalization of the charge Drude weight to first order in the strength of e-e interactions. The diagram in Fig.~\ref{fig:SM_five}a) gives
\begin{eqnarray} \label{eq:SM_chi_VC_def}
\chi_{hh}^{({\rm VC})}(q,i\omega_m) &=& N_{\rm f} (k_{\rm B} T)^2 \sum_{{\bm k},{\bm k}'} \sum_{\lambda,\lambda'} \sum_{\mu,\mu'} \sum_{\varepsilon_{n},\varepsilon_{n'}} v_{{\bm k}-{\bm k}'} 
\nonumber\\
&\times&
G_{\lambda}({\bm k}_-,i\varepsilon_n) G_{\lambda'}({\bm k}_+,i\varepsilon_n+i\omega_m)
\nonumber\\
&\times&
G_{\mu'}({\bm k}'_+,i\varepsilon_{n'}+i\omega_m) G_{\mu}({\bm k}'_-,i\varepsilon_{n'})
\nonumber\\
&\times&
{\cal D}_{\lambda\lambda'}({\bm k}_-,{\bm k}_+)
{\cal D}_{\lambda'\mu'}({\bm k}_+,{\bm k}'_+)
\nonumber\\
&\times&
{\cal D}_{\mu'\mu}({\bm k}'_+,{\bm k}'_-)
{\cal D}_{\mu\lambda}({\bm k}'_-,{\bm k}_-)
\nonumber\\
&\times&
\frac{\xi_{{\bm k}_-,\lambda} + \xi_{{\bm k}_+,\lambda'}}{2}
\frac{\xi_{{\bm k}'_-,\mu} + \xi_{{\bm k}'_+,\mu'}}{2}
~,
\end{eqnarray}
where ${\bm k}_\pm = {\bm k} \pm {\bm q}/2$, ${\bm k}'_\pm = {\bm k}' \pm {\bm q}/2$ and $\xi_{{\bm k},\lambda} = \lambda v_{\rm F} k - \mu$ is the energy measured from the chemical potential. At the same time the sum of the two self-energy diagrams in Fig.~\ref{fig:SM_five}b)-c) reads
\begin{eqnarray} \label{chi_SE_def}
\chi_{hh}^{({\rm SE})}(q,i\omega_m) &=& 2 N_{\rm f} (k_{\rm B} T)^2 \sum_{{\bm k},{\bm k}'} \sum_{\lambda,\lambda'} \sum_{\mu,\mu'} \sum_{\varepsilon_{n},\varepsilon_{n'}} v_{{\bm k}-{\bm k}'} 
\nonumber\\
&\times&
G_{\lambda}({\bm k}_-,i\varepsilon_n) G_{\lambda'}({\bm k}_+,i\varepsilon_n+i\omega_m)
\nonumber\\
&\times&
G_{\mu'}({\bm k}'_+,i\varepsilon_{n'}+i\omega_m)
\nonumber\\
&\times&
G_{\mu}({\bm k}_+,i\varepsilon_{n}+i\omega_m)
\nonumber\\
&\times&
{\cal D}_{\lambda\lambda'}({\bm k}_-,{\bm k}_+)
{\cal D}_{\lambda'\mu'}({\bm k}_+,{\bm k}'_+)
\nonumber\\
&\times&
{\cal D}_{\mu'\mu}({\bm k}'_+,{\bm k}_+)
{\cal D}_{\mu\lambda}({\bm k}_+,{\bm k}_-)
\nonumber\\
&\times&
\left(\frac{\xi_{{\bm k}_-,\lambda} + \xi_{{\bm k}_+,\lambda'}}{2}\right)^2
~.
\end{eqnarray}
Performing the Matsubara sums over $\varepsilon_n$ and $\varepsilon_{n'}$ and expanding Eqs.~(\ref{eq:SM_chi_VC_def}) and~(\ref{chi_SE_def}) to ${\cal O}(q^2/\omega^2)$ we get
\begin{eqnarray} \label{eq:SM_chi_VC_exp}
\chi_{hh}^{({\rm VC})}(q,i\omega_m) &\to& -\frac{N_{\rm f} q^2}{2 \omega^2} \sum_{{\bm k},{\bm k}'} v_{{\bm k}-{\bm k}'} n_{\rm F}'(\xi_{{\bm k},+}) n_{\rm F}'(\xi_{{\bm k}',+}) 
\nonumber\\
&\times&
\cos(\varphi_{\bm k})
\cos(\varphi_{{\bm k}'})
\big[1+\cos(\varphi_{\bm k}-\varphi_{{\bm k}'}) \big]
\nonumber\\
&\times&
\xi_{{\bm k},+}\xi_{{\bm k}',+}
~,
\end{eqnarray}
and
\begin{eqnarray} \label{eq:SM_chi_SE_exp}
\chi_{hh}^{({\rm SE})}(q,\omega) &=& \chi_{hh}^{({\rm SE}-1)}(q,\omega) + \chi_{hh}^{({\rm SE}-2)}(q,\omega)
\nonumber\\
&\to&
\frac{N_{\rm f} q^2}{2 \omega^2} \sum_{{\bm k},{\bm k}'} v_{{\bm k}-{\bm k}'} n_{\rm F}'(\xi_{{\bm k},+}) n_{\rm F}'(\xi_{{\bm k}',+}) \xi_{{\bm k},+}^2
\nonumber\\
&\times&
\cos(\varphi_{\bm k})
\cos(\varphi_{{\bm k}'})
\big[1+\cos(\varphi_{\bm k}-\varphi_{{\bm k}'}) \big]
\nonumber\\
&-&
\frac{N_{\rm f} q^2}{2 \omega^2} \sum_{{\bm k},{\bm k}'} v_{{\bm k}-{\bm k}'} n_{\rm F}'(\xi_{{\bm k},+}) \big[1-n_{\rm F}(\xi_{{\bm k}',+})\big]
\nonumber\\
&\times&
\xi_{{\bm k},+}^2
\cos(\varphi_{\bm k})
\sin(\varphi_{\bm k}-\varphi_{{\bm k}'})
\nonumber\\
&\times&
\left[\frac{\sin(\varphi_{\bm k})}{k} - \frac{\sin(\varphi_{{\bm k}'})}{k'} \right]
~.
\end{eqnarray}
Note that in the case of the number-density response function we obtain very similar expressions. The differences between the energy-density and number-density linear response functions reside in the extra factors $\xi_{{\bm k},+} \xi_{{\bm k}',+}$ and $\xi_{{\bm k},+}^2$ on the right-hand sides of Eqs.~(\ref{eq:SM_chi_VC_exp}) and~(\ref{eq:SM_chi_SE_exp}), respectively. In the case of the number-density response function these factors are absent. As we show below, the vertex correction to the energy-density response function due to the diagram in Fig.~\ref{fig:SM_five}a) vanishes because of the extra factor $\xi_{{\bm k},+} \xi_{{\bm k}',+}$.

We consider first Eq.~(\ref{eq:SM_chi_VC_exp}). Note that $n_{\rm F}'(\xi_{{\bm k},+})$ and $n_{\rm F}'(\xi_{{\bm k}',+})$ constrain $\xi_{{\bm k},+} \sim 0$ and $\xi_{{\bm k}',+}\sim 0$. This in turn implies that $k=k_{\rm F}$ and $k'=k_{\rm F}$, and that $v_{{\bm k}-{\bm k}'}$ is a function only of the difference of the angles $\varphi_{\bm k}-\varphi_{{\bm k}'}$. The sums over the moduli of ${\bm k}$ and ${\bm k}'$ are thus proportional to
\begin{eqnarray} \label{eq:SM_I_VC}
{\cal I}_{\rm VC} &=& T^2 \int_{-\mu/T}^{+\infty} d{\bar \varepsilon}~ {\bar \varepsilon} n_{\rm F}'({\bar \varepsilon}) \int_{-\mu/T}^{+\infty} d{\bar \varepsilon}'~ {\bar \varepsilon}' n_{\rm F}'({\bar \varepsilon}') 
\nonumber\\
&=&
{\cal O}\left(\frac{T^4}{\mu^2}\right)
~.
\end{eqnarray}
Hereafter we define $n_{\rm F}(x) = (e^x+1)^{-1}$. The first term on the right-hand side of Eq.~(\ref{eq:SM_chi_SE_exp}) is very similar to the case just discussed. However, in this case the sums over the moduli of ${\bm k}$ and ${\bm k}'$ are proportional to
\begin{eqnarray} \label{eq:SM_I_SE_1}
{\cal I}_{{\rm SE}-1} &=& T^2 \int_{-\mu/T}^{+\infty} d{\bar \varepsilon}~ {\bar \varepsilon}^2 n_{\rm F}'({\bar \varepsilon}) \int_{-\mu/T}^{+\infty} d{\bar \varepsilon}'~ n_{\rm F}'({\bar \varepsilon}') 
\nonumber\\
&=&
\frac{\pi^2 (k_{\rm B} T)^2}{3} + {\cal O}\left(\frac{T^4}{\mu^2}\right)
~.
\end{eqnarray}
Thus $I_{\rm VC}$ is negligible compared to $I_{{\rm SE}-1}$. Note that, in the case of the number-density response function, the vertex diagram and the contribution of the self-energy diagram analogous to Eq.~(\ref{eq:SM_I_SE_1}) were both different from zero and canceled against each other. In the case of the energy-density response function the vertex correction is instead negligible in the limit $T\to 0$ as compared to the self-energy correction. Finally, let us consider the second term on the right-hand side of Eq.~(\ref{eq:SM_chi_SE_exp}). This time only $k$ is bounded to the Fermi surface, and thus $v_{{\bm k}-{\bm k}'}$ is a function of both the angle $\varphi_{\bm k}-\varphi_{{\bm k}'}$ and the modulus $k'$. The integral over the modulus of ${\bm k}$ is proportional to
\begin{eqnarray} \label{eq:SM_I_SE_2}
{\cal I}_{{\rm SE}-2} &=& T^2 \int_{-\mu/T}^{+\infty} d{\bar \varepsilon}~ {\bar \varepsilon}^2 n_{\rm F}'({\bar \varepsilon}) 
\nonumber\\
&=&
\frac{\pi^2 (k_{\rm B} T)^2}{3} + {\cal O}\left(\frac{T^4}{\mu^2}\right)
~.
\end{eqnarray}
Finally, we get
\begin{eqnarray} \label{eq:SM_chi_hh_final}
\chi_{hh}^{({\rm VC})}(q,\omega) &=& {\cal O}\left(\frac{T^4}{\mu^2}\right)
~,
\nonumber\\
\chi_{hh}^{({\rm SE})}(q,\omega) &=& \frac{\pi^2 (k_{\rm B} T)^2}{3} \chi_{nn}^{({\rm SE})}(q,\omega) + {\cal O}\left(\frac{T^4}{\mu^2}\right)
~,
\nonumber\\
\end{eqnarray}
where $\chi_{nn}^{({\rm SE})}(q,\omega)$ is the self-energy contribution to the number-density response function up to ${\cal O}(q^2/\omega^2)$ and to first order in the strength of electron-electron interactions. Eq.~(\ref{eq:SM_chi_hh_final}) implies that, contrary to the charge Drude weight, the thermal Drude weight contains only self-energy contributions and no vertex correction. This confirms the result we found with the Landau's phenomenological theory of normal Fermi liquids, namely that the thermal Drude weight is proportional to the renormalized Fermi velocity $v_{\rm F}^\star$, while the charge Drude weight is proportional to the product $v_{\rm F}^\star(1+F_1^{\rm s})$. Indeed, the Landau parameters~\cite{Giuliani_and_Vignale} $F_n^{\rm s}$ encode  the vertex corrections to the properties of the system.


\begin{thebibliography}{77}
\bibitem{entropy_density}
  The entropy of quasiparticles is defined as $S = -\sum_{\bm k} \big[ n_{\bm k} \ln(n_{\bm k}) + (1-n_{\bm k}) \ln(1-n_{\bm k}) \big]$, where $n_{\bm k} \equiv n_{\bm k}({\bm r},t)$ was defined in the main text. Expanding to first order in $n_{1,{\bm k}}({\bm r},t)$ we get
  %
  \begin{eqnarray}
  \delta S &\equiv& S-S_0 = -\sum_{\bm k} n_{1,{\bm k}}({\bm r},t) \ln \left[ \frac{n_0(\xi_{{\bm k}}^\star)}{1-n_0(\xi_{{\bm k}}^\star)} \right]
  \nonumber\\
  &=&
  \sum_{\bm k} \frac{\xi_{\bm k}^\star}{k_{\rm B} T} n_{1,{\bm k}}({\bm r},t)
  ~.
  \end{eqnarray}
  %
  Here $S_0 = -\sum_{\bm k} \big\{ n_0(\xi_{\bm k}^\star) \ln[n_0(\xi_{\bm k}^\star)] + [1-n_0(\xi_{\bm k}^\star)] \ln[1-n_0(\xi_{\bm k}^\star)] \big\}$ is the entropy of quasiparticles at equilibrium.
\end{thebibliography}
\end{document}